\documentclass[aps,prd,onecolumn,eqsecnum,amsmath,nofootinbib,preprintnumbers]{revtex4}%

\usepackage{color,graphicx,float,subfigure}
\usepackage{amsfonts,amssymb,theorem,mathrsfs,times}
\usepackage{bm}
\usepackage{mathtools}
\usepackage{amsfonts,amssymb,theorem,mathrsfs}
\usepackage{dsfont}
\usepackage{setspace}
\usepackage{graphicx}
\usepackage{ulem}
\usepackage[colorlinks,linkcolor=blue,anchorcolor=blue,citecolor=blue]{hyperref}
\usepackage{cancel}
{\theorembodyfont{\upshape}
}
{\theorembodyfont{\upshape}
}
{\theorembodyfont{\upshape}
}
{\theorembodyfont{\upshape}
}
{\theorembodyfont{\upshape}
}
{\theorembodyfont{\upshape}
}

\newcommand{\dalm}{\kern1pt\vbox{\hrule height 0.9pt\hbox{\vrule width
0.9pt\hskip 2.5pt\vbox{\vskip 5.5pt}\hskip 3pt\vrule width
0.3pt}\hrule height 0.3pt}\kern1pt}

\begin{document}
\preprint{\hfill {\small{ICTS-USTC/PCFT-23-39}}}
\title{The appearance of the regular black hole with a stable inner horizon}

%

\author{ Li-Ming Cao$^{a\, ,b}$\footnote{e-mail
address: caolm@ustc.edu.cn}}

\author{ Long-Yue Li$^b$\footnote{e-mail
address: lily26@mail.ustc.edu.cn}}

\author{ Xia-Yuan Liu$^b$ \footnote{e-mail
address: liuxiayuan@mail.ustc.edu.cn }}

\author{ Yu-Sen Zhou$^b$\footnote{e-mail
address: zhou\_ys@mail.ustc.edu.cn}}

\affiliation{$^a$Peng Huanwu Center for Fundamental Theory, Hefei, Anhui 230026, China}

\affiliation{${}^b$
Interdisciplinary Center for Theoretical Study and Department of Modern Physics,\\
University of Science and Technology of China, Hefei, Anhui 230026,
China}


\date{\today}


\begin{abstract}
The strong cosmic censorship conjecture, which states that the evolution of generic initial data will always produce a globally hyperbolic spacetime, is hard to be tested by astronomical observations.
In this paper, we study the appearance of the regular black hole without mass inflation, which violates the strong cosmic censorship conjecture.
Since the inner horizon is stable, the photons entering the two horizons of the regular black hole in the preceding companion universe can come out from the white hole in our universe.
These rays create a novel multi-ring structure, which is significantly different from the image of the Schwarzschild black hole.
This serves a potential method to test the strong cosmic censorship conjecture.
\end{abstract}


\maketitle


\section{Introduction}
It has been years since the images of the supermassive objects at the center of M87 \cite{EventHorizonTelescope:2019dse} and Milky Way galaxies \cite{EventHorizonTelescope:2022wkp} was captured by the Event Horizon Telescope (EHT).
These provide the evidence for the existence of black holes.
In these images, the dark area is surrounded by a bright disk-like structure, which is illuminated dominantly by the direct emission of the accretion disk.
In theoretical calculations, there is a narrow bright ring besides the direct emission from the accretion disk \cite{Gralla:2019xty}.
This is so called lensing ring, which is constituted by the ray emitted near the photon sphere.
The photon sphere is a region of spacetime where photons can orbit a black hole an infinite number of times. In the image plane, the curve corresponding to the photon sphere is referred to as the critical curve. Surrounding this critical curve in the image plane, there is a region of enhanced brightness. Within this region, two distinct phenomena occur: the photon ring and the lensing ring. The photon ring is formed by light rays that intersect the accretion disk more than twice, while the lensing ring is formed by light rays that intersect the accretion disk exactly twice.
However, the photon ring is so close to the lensing ring that it can not be distinguished, and its contribution to the overall intensity is negligible.
Although the critical curve and the photon ring are fully determined by the background geometry, the optical appearance of a compact object is heavily influenced by the position and profile of light source.
This gives us a window through which we can get the background geometry and accretion disk information from the black hole image.

Last century, Hawking and Penrose demonstrated that singularities occur unavoidably in general relativity (GR) \cite{Penrose:1964wq,Hawking:1970zqf}.
Weak cosmic censorship conjecture (WCCC) \cite{Penrose:1969pc} postulate that singularities, where physical laws break down, always hide behind the horizon.
Furthermore, the presence of the singularity, i.e., the existence of incomplete geodesics, threatens the predictability of GR.
Therefore, regular black holes without singularities, which are constructed by relaxing some conditions in the singularity theorem, attract much attention.
Following Bardeen's work \cite{Bardeen}, many regular black holes were discovered, such as the loop black hole \cite{Modesto:2008im}, the Hayward black hole \cite{Hayward:2005gi}, the quantum-corrected black hole \cite{Lewandowski:2022zce} and so on \cite{Bambi:2023try,Lan:2023cvz}.
Regular black holes are usually constructed by replacing the central singularity with a non-singular core.
They are classified as de Sitter core and Minkowski core according to the asymptotic behavior near  $r=0$.
The WCCC is discussed in the rotate regular black hole in \cite{Ghosh:2022gka}.
Even though regular black holes avoid singularities at their centers, most of them possess an unstable inner horizon.
In fact, all spherically symmetric regular black holes have inner horizons \cite{Carballo-Rubio:2019fnb}.
If the Cauchy horizon is stable, the Penrose diagram of the spacetime can be extended to consist of infinite repeated universes, thereby implying an infinite series of discrete inner horizons.
The Cauchy horizon of a certain initial data surface is just a part of certain inner horizon.
If we confine ourselves to a single universe, the inner horizon can be identical to the Cauchy horizon.
Hereafter, we will mainly use the terminology ``inner horizon" instead of ``Cauchy horizon" since their subtle distinctions won't be crucial for our discussion.
The extension beyond Cauchy horizon will break down the predictability of GR, and the classical determinism is lost.
Fortunately, most of these inner horizons of regular black holes are unstable, which are caused by  mass inflation and will convert into  null singularities \cite{Brown:2011tv,Iofa:2022dnc,Carballo-Rubio:2021bpr}.
The instability of the inner horizon prevents the spacetime extending beyond it, and thus avoid the problem of predictability.
This picture supports the strong cosmic censorship conjecture (SCCC), which states that a physical spactime is always globally hyperbolic \cite{Penrose:1978,Penrose:1979azm}.
In the modern language of partial differential equations, the evolution of generic initial data will always produce a globally hyperbolic spacetime.
In other words, a stable Cauchy horizon is in contradiction with the SCCC.
Examination of the SCCC is an important topic in mathematical physics nowadays \cite{Dafermos:2002ka,Dafermos:2003vim,Dafermos:2003wr,Dafermos:2012np,Dafermos:2017dbw}.
However, it is challenging to verify this conjecture through astronomical observations due to the lack of proper ways.

The image of the regular black hole has been studied in recent years \cite{Guo:2021wid,DeMartino:2023ovj,Tsukamoto:2017fxq,Tsukamoto:2014tja,Jiang:2023img,Ghosh:2022gka}.
Their shapes are similar to the case of the Schwarzschild black hole or the Kerr black hole, except for some differences in intensity, position of the photon ring and so on.
Because the images of the regular black holes in \cite{Guo:2021wid,DeMartino:2023ovj} and the Schwarzschild black hole are mainly determined by the direct emission, which is strongly influenced by the accretion disk, it is not easy to distinguish the images of regular black holes with the Schwarzschild one.
It is exciting that the quantum-corrected black hole may have a novel image \cite{Zhang:2023okw}.
This is a regular black hole proposed by the quantum Oppenheimer-Snyder and Swiss Cheese models in the framework of Loop Quantum Gravity (LQG) \cite{Lewandowski:2022zce}.
The photon can enter the event horizon of a companion black hole in the preceding universe, come out from the white hole in our universe and finally be captured by the observer.
These rays produces many new bright rings in the image of the black hole, especially some of these rings are distinctly visible within the shadow.
This multi-ring structure may be detected astronomically.
However, it has been proved that the inner horizon of the quantum-corrected black hole is unstable in \cite{Cao:2023aco}.
After a small perturbation, the mass inflation occurs and the inner horizon becomes a null singularity.
This prevents the spacetime to extend beyond the inner horizon, and the photons can not pass though it.
Therefore, whether this novel multi-ring structure will occur is still unclear.

Intriguingly, a regular black hole with a stable inner horizon was proposed in \cite{Carballo-Rubio:2022kad}.
The vanishing surface gravity of the inner horizon eliminates the possibility of the mass inflation instability.
Using the ray-tracing method \cite{Gralla:2019xty}, we draw its image in section \ref{S3}.
Similar to the quantum-corrected black hole, which ignoring the instability of the inner horizon, a multi-ring structure emerges.
The stable inner horizon ensures that the photons can pass though the two horizons in the companion black hole in the preceding universe, and come out from the white hole in our universe.
Therefore, unlike the quantum-corrected black hole, this multi-ring structure is physically permissible.
However, as we mentioned above, the stable Cauchy horizon will violate the SCCC.
Hence, the presences of multi-ring structure in observation may suggest that SCCC is violated.
This provides us a way to test the SCCC in astronomical observations.

However, the multi-ring structure also occurs in the compact object and the wormhole \cite{Olmo:2021piq,Guerrero:2022qkh,Guerrero:2022msp,Olmo:2023lil}.
What they have in common is that they all have no horizon.
In \cite{Guerrero:2022qkh}, the generalized black bounce-type geometry and its image was studied.
As the radius of the throat of the wormhole decreases, the horizons appear and the spacetime turns into a black hole.
The metric and the  effective potential of the black hole are similar to that of the wormhole, while the multi-ring structure only appears in the image of the wormhole.
This is because the existence of horizon prevents the escape of the photons which have fallen into the horizon.
In our investigation, the photons from the companion black hole in the preceding universe can effect the image photographed by the observer in our universe.
Therefore, our result also suggests that the multi-ring structure can appear in the object with horizon.
To distinguish the compact object, wormhole and the regular black hole, we require additional observation, such as the experiment involving a significant redshift in the presence of an event horizon.

This paper is organized as follows.
In section \ref{S2}, we introduce the regular black hole with a stable inner horizon and get its Penrose diagram.
We then consider thin disk emission at different locations and investigate their observational appearance in section \ref{S3}.
In section \ref{S4}, the image of the regular black with different parameters is investigated.
Finally, we give conclusions and discussions in section \ref{S5}.

\section{The regular black hole with a stable inner horizon}\label{S2}
\begin{figure}[htbp]
\centering
\subfigure[]{
\includegraphics[width=8cm]{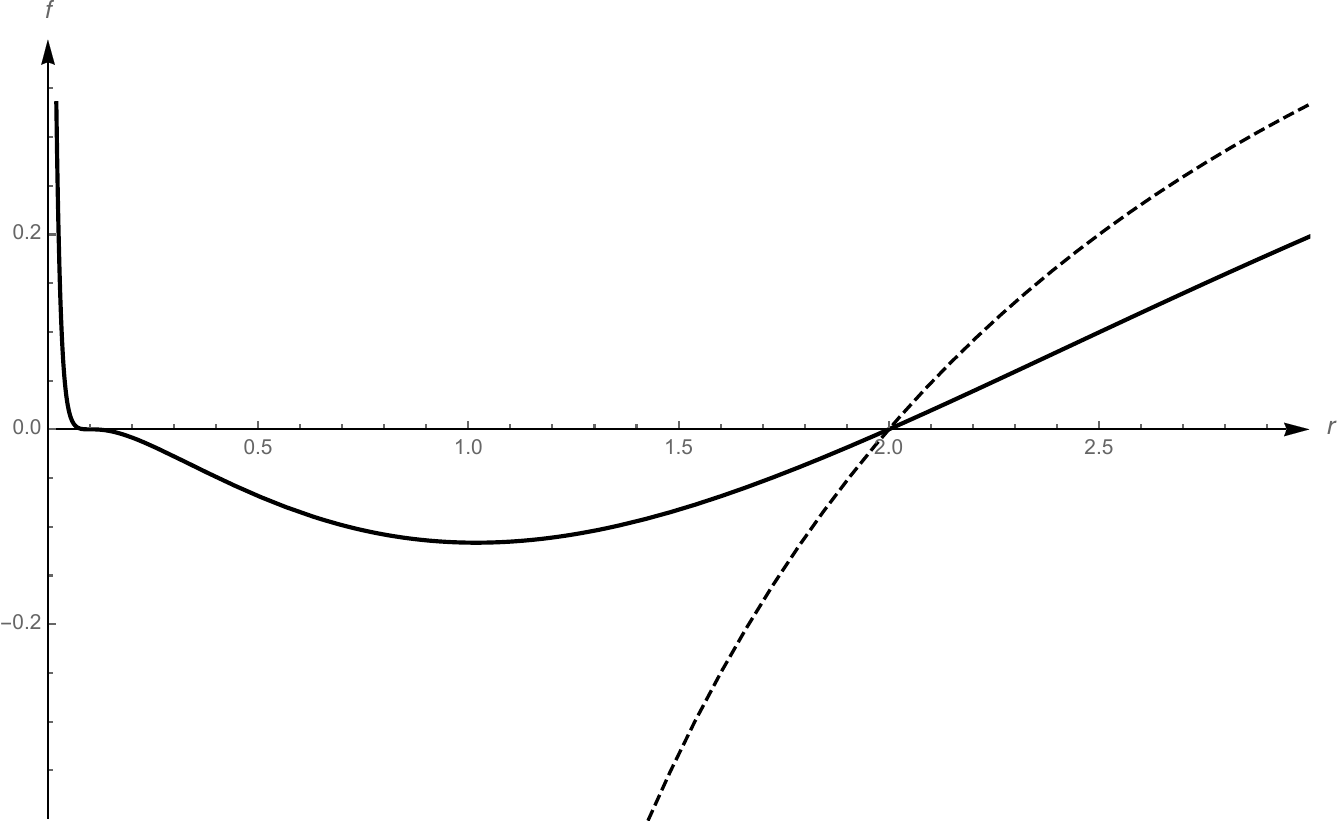}
\label{f}}
\hspace{1cm}
\subfigure[]{
\includegraphics[width=8cm]{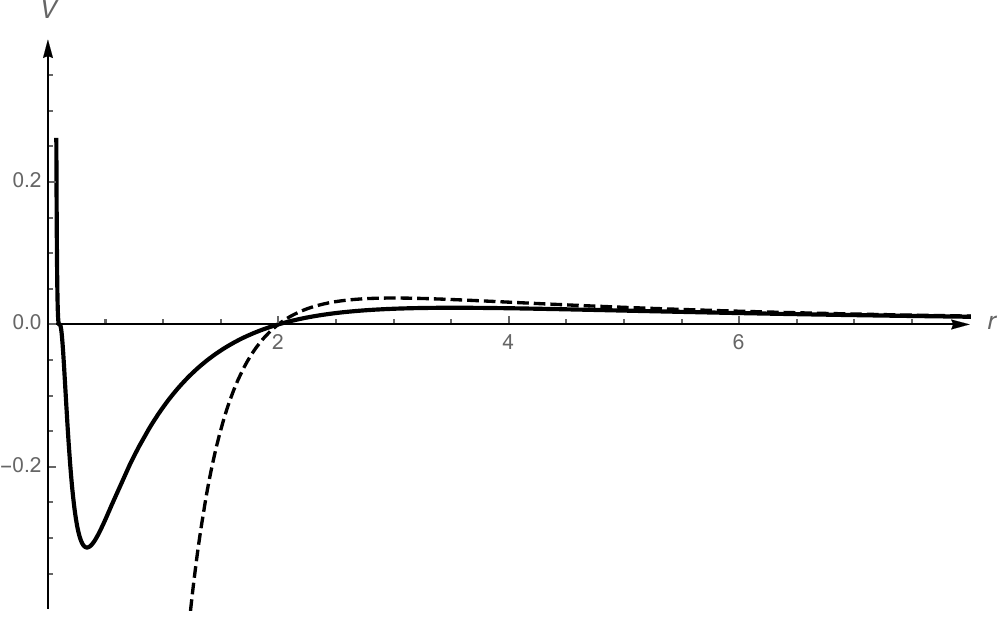}
\label{V}}
\caption{The metric function $f(r)$ (a) and the effective potential (b) of the regular black hole (solid curve) with $M=1,r_-=0.1,r_+=2,b_2=5$ and the Schwarzschild black hole (dashed curve) with $M=1$. }

\end{figure}

A regular black hole with a stable inner horizon is proposed in \cite{Carballo-Rubio:2022kad}. In Schwarzschild-like coordinates, the metric of this spherically symmetric regular black hole can be written as
\begin{eqnarray}
\mathrm{d}s^2=-f(r)\mathrm{d}t^2+\frac{1}{f(r)}\mathrm{d}r^2+r^2\mathrm{d}\theta^2+r^2\sin^2\theta\mathrm{d}\phi^2  \,,
\end{eqnarray}
where
\begin{eqnarray}
f(r)=\frac{(r-r_-)^3(r-r_+)}{(r-r_-)^3(r-r_+)+2Mr^3+b_2r^2}  \,,
\end{eqnarray}
$M$ is the mass parameter of the black hole, $r_+$ and $r_-$ are the radii of the event horizon and the inner horizon of the black hole respectively.
The parameter $b_2$ is sufficiently large and is related to the parameter $a_2$ as follows
\begin{eqnarray}
b_2=a_2-3r_-(r_++r_-)  \,.
\end{eqnarray}
In the case that $a_2\gtrsim 9r_+r_-/4$, $r_-\ll2M$, and $r_+\approx2M$, the metric outside the event horizon is similar to that of the Schwarzschild one.
Furthermore, when $b_2=r_-=0$ and $r_+=2M$, the metric exactly reduces to that of the Schwarzschild black hole. The metric function $f(r)$ is plotted in Fig.\ref{f}.

The Ricci scalar and the Kretschmann scalar are finite for arbitrary $r$, and thus the black hole has no curvature singularity.
In the asymptotic limit $r\rightarrow\infty$, it also tends to the Schwarzschild solution.
However, in the limit $r\rightarrow0$, it embodies a de Sitter core-type behavior
\begin{eqnarray}
f \rightarrow 1-\frac{b_2}{r_-^3 r_+}r^2 \,.
\end{eqnarray}
It is worth noting that the surface gravity of the inner horizon vanishes, i.e., $\kappa_-=0$, while that of the event horizon is nonzero, i.e., $\kappa_+\neq 0$.
Therefore, the observer will receive a finite flux near the inner horizon after a small perturbation \cite{1982RSPSA.384..301C}.
Besides, based on the model by Ori  \cite{Ori:1991zz} and the general DTR relation~\cite{Brown:2011tv}, it can be shown that the inner horizon is stable~\cite{Carballo-Rubio:2022kad}.

\begin{figure}[htbp]
\centering

\subfigure[]{\includegraphics[width=2cm]{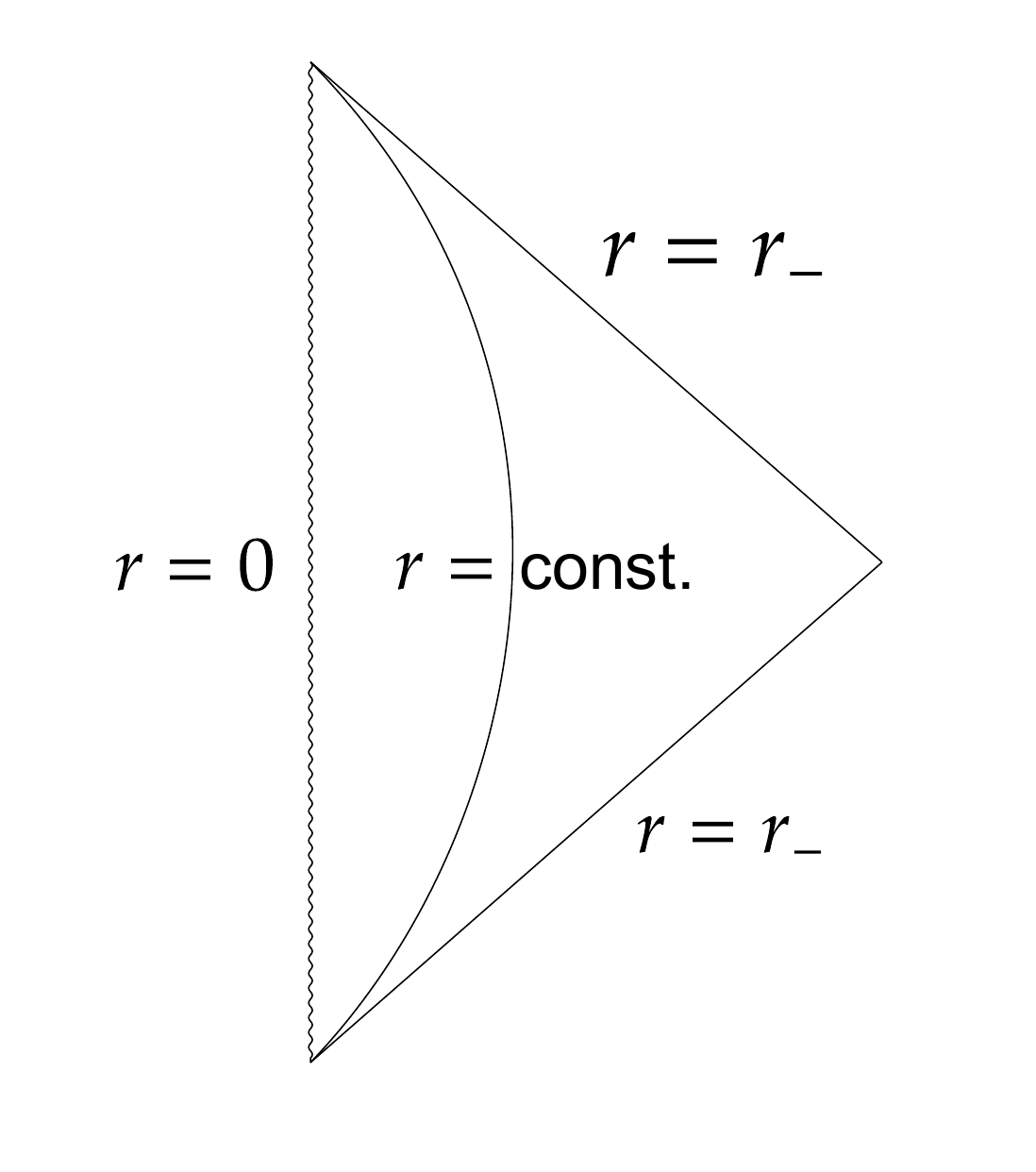}
\label{r=0}}
\hspace{1cm}
\subfigure[]{\includegraphics[width=5cm]{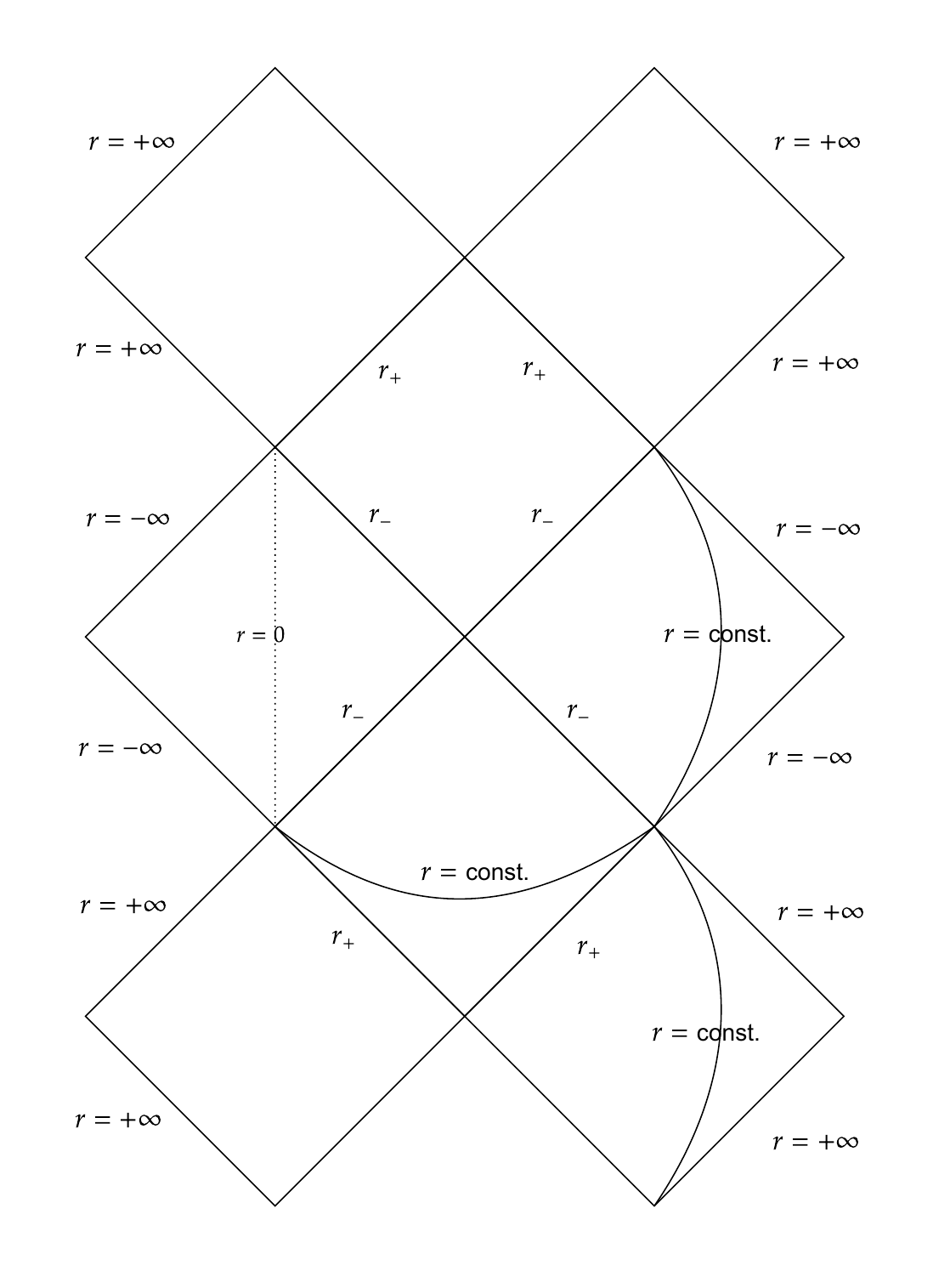}\label{penrose}}
\caption{(a) is the block with $r<r_-$ of the RN black hole and (b) is the Penrose diagram of the regular black hole. $r=0$ of the regular black hole is not a singularity.}
\end{figure}

In order to investigate the shadow of the regular black hole, it is crucial to understand the trajectory of photons.
Therefore, it is nature to study the global casual structure of the black hole by drawing the Penrose diagram.
To draw the Penrose diagram, usually one needs to find the Kruska-type coordinate.
But it turns out to be quite difficult and is not necessary \cite{Carter:2009nex,Walker:1970}.
As described in \cite{Walker:1970}, the Penrose diagram is made up of some blocks that are glued together according to some given rules.
The blocks of the regular black hole with $r>r_+$ and $r_-<r<r_+$ are quite similar to that of the RN black hole.
But things become different in the case of the block with $r<r_-$.
For RN black hole, $r=0$ is the singularity, thus the block with $r<r_-$ is a triangle, like Fig.\ref{r=0}, and the geodesic end at $r=0$.
As for the regular black hole, $r=0$ is not a singularity.
As a consequence, the geodesic can not terminate at $r=0$ and must be extended.
Therefore, the block with $r<r_-$ of the regular black hole is not a triangle.

There are two ways to extend the geodesic at $r=0$ in this block.
One is identifying $r=0$ with that of an identical universe.
This is similar to the case of the Kerr black hole and thus the Penrose diagram is similar to that of the Kerr black hole, too.
Another way is to extend the coordinate $r$ to $(-\infty,0)$.
In this case, the region $r\in(-\infty,r_-)$ is asymptotically flat and regular.
The geodesic can be extended to $r\rightarrow-\infty$.
Both two ways above result in the same Penrose diagram, as shown in Fig.\ref{penrose}.

The motion of a massless particle with energy $E$ and angular momentum $L$ is governed by the equation
\begin{eqnarray}
\dot{r}^2+V(r)=\frac{1}{b^2}  \,,
\end{eqnarray}
where
\begin{eqnarray}
V(r)=\frac{f(r)}{r^2}
\end{eqnarray}
is the effective potential, and $b=L/E$ is the impact parameter, and $``\cdot"$ denotes the derivative with respect to the so-called affine parameter.
The effective potential $V(r)$ is depicted in Fig.\ref{V}.
As can be seen here, the effective potential $V(r)$ of a regular black hole and that of a Schwarzschild black hole is very different, especially near $r=0$.
When $r\rightarrow0$, the the effective potential $V(r)\rightarrow-\infty$ for the Schwarzschild black hole, while $V(r)\rightarrow+\infty$ for the regular black hole.
This significant distinction leads to the difference in the appearance between the two black holes.

The coordinate of the critical curve, $r_m$, is determined by
\begin{eqnarray}
V(r_m)=\frac{1}{b_c^2} \,, \quad V'(r_m)=0 \,, \quad V''(r_m)<0  \,,
\end{eqnarray}
where the photons will surround the black hole in the circular orbit for infinite times, and $b_c$ is the corresponding impact parameter.
As shown in Fig.\ref{V}, there is only one critical curve for a  regular black hole. Obviously, the photon with non-zero angular momentum can not reach $r=0$ and it will bounce back at a certain minimum value of $r$.

\section{The ray-tracing method and the appearance of the regular black hole}\label{S3}
In this section, we give a brief introduction to the ray-tracing method  \cite{Gralla:2019xty} and study the appearance of the regular black hole.

\begin{figure}[htb]
  \centering
  \includegraphics[width=7cm]{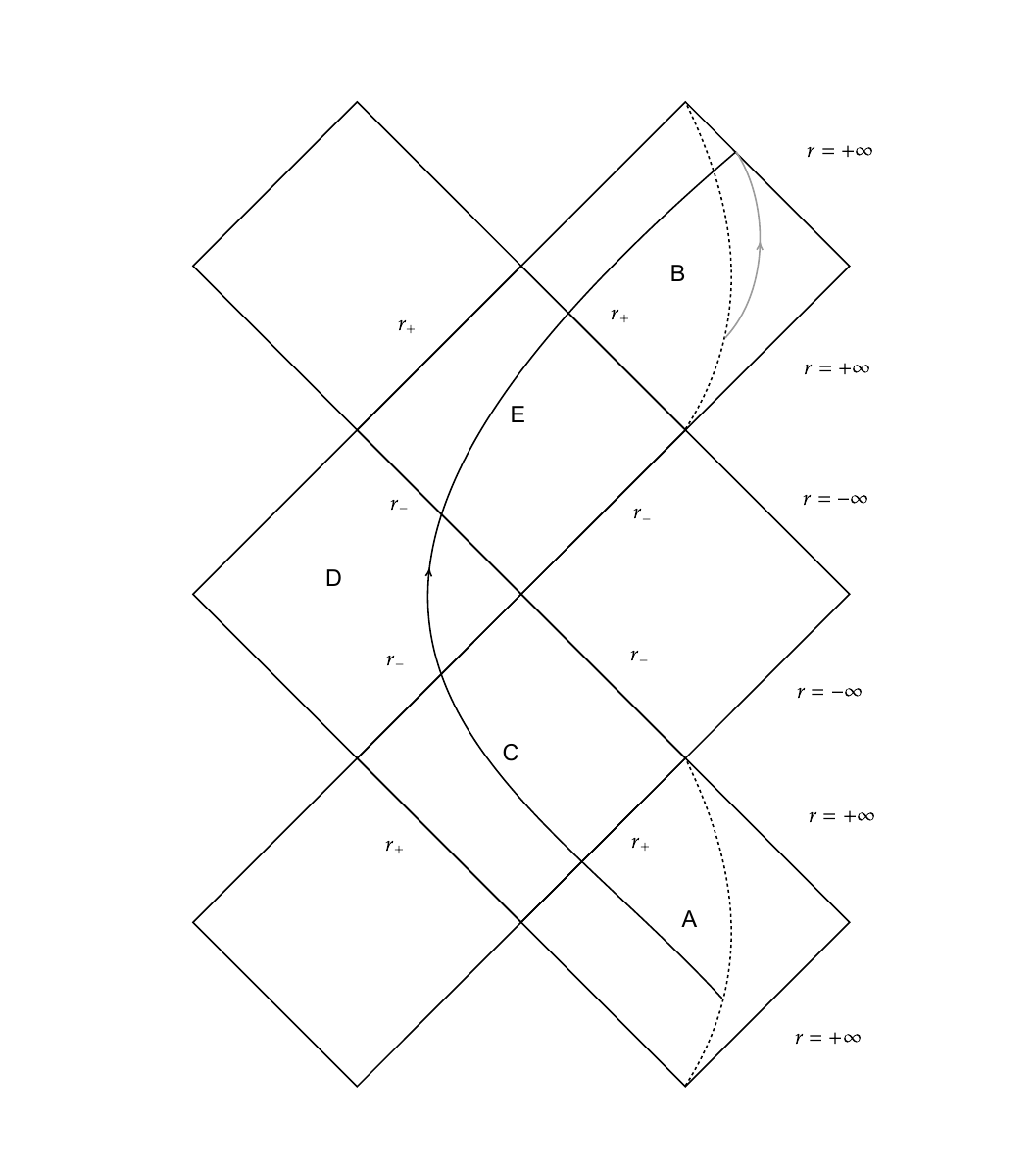}
  \caption{The photon trajectory in the Penrose diagram. The two dashed curve in universe $A$ and $B$ is the world lines  of the photon sphere. The gray curve is the photon trajectory from the photon sphere to the observer in the same universe $B$. And the black curve is the photon trajectory from the photon sphere in $A$ to the observer in $B$.}
  \label{geo}
\end{figure}

Consider an observer in universe $B$, positioned near the null infinity to receive the photons emitted from the region near the event horizon.
The process is depicted in the Penrose diagram shown in Fig.\ref{geo}.
The accretion disk is around the black hole (or white hole) and is sharply peaked at the photon sphere, which is also called ``photon sphere" where the light orbits around it infinite times.
In Fig.\ref{geo} the photon sphere is described by the dashed curve.
As we can see, the observer in universe $B$ can receive the photons from two sources. One is from the accretion disk in universe $B$, illustrated as the gray curve in Fig.\ref{geo}.
Another is showed as the black curve in Fig.\ref{geo}.
The photons are emitted from the the accretion disk in universe $A$, subsequently falling into the event horizon, then traverse the inner horizon, and arrive at the universe $D$.
Then they travel through the inner horizon and the event horizon of the white hole in turn, and are finally captured by the observer in universe $B$.
These two kinds of source of photons lead to different images.
Therefore, we will discuss these two cases separately.

\begin{figure}[htb]
  \centering
  \includegraphics[width=6cm]{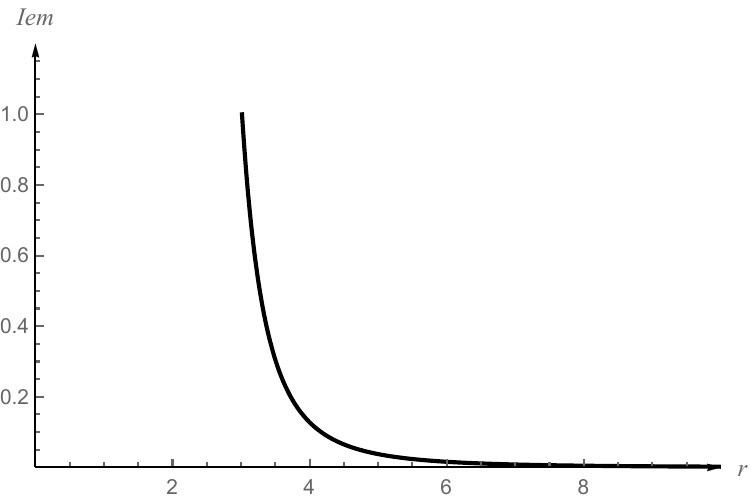}
  \caption{The emission profiles with $r_m=3.014$. Here $r_m$ is the photon sphere of the regular black hole with $r_+=2, r_-=1/50, M=1, b_2=0$.}
  \label{Iem}
\end{figure}

Each of intersections of light with the accretion disk contributes to the intensity received by the observer near null infinity.
Besides, considering the effect of gravitational redshift on the intensity of the emission, the intensity of the light received by the observer, $I_{obs}$, is \cite{Gralla:2019xty,DeMartino:2023ovj,Rybicki:2004hfl}
\begin{eqnarray}
I_{obs}=\sum_n I_{em}(r)f^2(r)|_{r=r_n}  \,,
\end{eqnarray}
where $r_n$ is the position of $n$-th intersection with the accretion disk.
We adopt the geometrically and optically thin accretion model, whose emission brightness is given by \cite{Guerrero:2022qkh}
\begin{eqnarray}
I_{em}(r) = \begin{cases}
\dfrac{1}{(r-r_m+1)^3} , & \text{if } r \geqslant r_m\, , \\
0, & \text{if } r<r_m\, .
\end{cases}
\end{eqnarray}
We assume the light emitted is monochromatic, so the photons are emitted with the same frequency. The emission is sharply peaked and ends at the photon sphere, as shown in Fig.\ref{Iem}

\subsection{The image of the black hole  formed by the photons emitted in universe $B$}
\begin{figure}[htb]
  \centering
  \includegraphics[width=6cm]{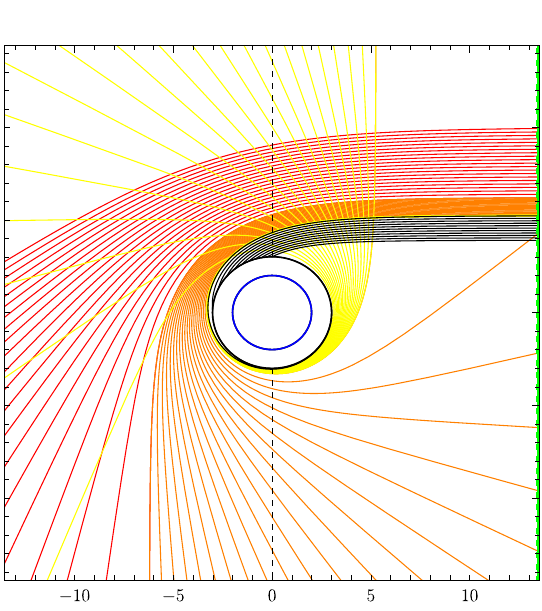}
  \caption{The trajectory of the photon around the regular black hole with $r_+=2, r_-=1/50, M=1, b_2=0$. The green line denotes the observation screen at the null infinity. The black circle corresponds to  the photon sphere and the blue circle corresponds to the event horizon. The dashed line is the accretion disk, which is the light source and located at the equatorial plane. The colored curves correspond to $n<3/4$ for red, $3/4<n<5/4$ for orange, $n>5/4$ for yellow, and $b<b_c$ for black.}
  \label{trajout}
\end{figure}

In this case, the null geodesic is depicted as the gray curve in Fig.\ref{geo}.
The entire trajectory of photons is in the universe $B$.
We have illustrated the trajectory of the photon around the regular black hole with $r_+=2, r_-=1/50, M=1, b_2=0$ in Fig.\ref{trajout} by using the ray-tracing method \cite{Gralla:2019xty}.

The normalized number of orbits $n=\phi/(2\pi)$ relates to the number of intersections with the equatorial plane of a particular light ray, where $\phi$ is the azimuthal angle.
The null geodesic with $b>6.215$ is shown as red curves with $n<3/4$. They intersect with the equatorial plane (dashed line) for one time, so is called ``direct emission".
The null geodesic with $5.264<b<6.215$ is colored by orange with $3/4<n<5/4$. They intersect with the equatorial plane for two times and are called ``lensing ring".
At each intersection, newly emitted photons from the accretion disk join in the journey towards the screen.
The null geodesic with $5.230<b<5.264$ is colored by yellow with $n>5/4$. They cross the equatorial plane more than 3 times and is called ``photon ring".
The impact parameter of the critical curve is $b_c=5.231$.
The rays with $b<b_c$ is direct emission, which is similar to the situation of $b>6.215$ and is shown as the black curves.
And the rays with $b<3.904$ can not intersect with the accretion disk and, therefore, do not contribute to the intensity.

\begin{figure}[htbp]
\centering

\subfigure[]{\includegraphics[width=8cm]{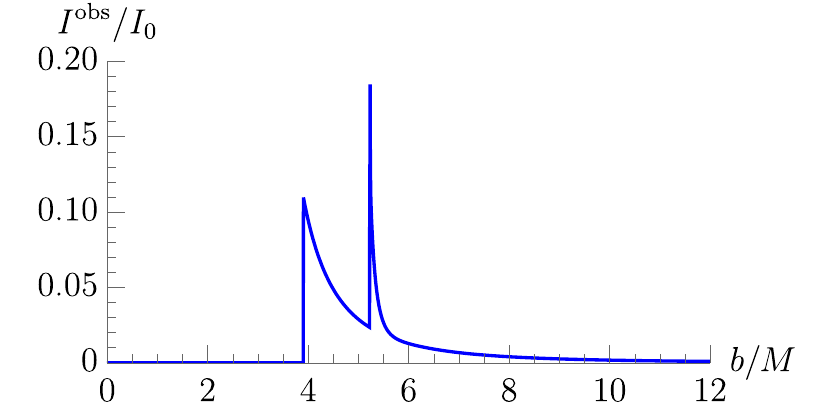}}
\subfigure[]{\includegraphics[width=5cm]{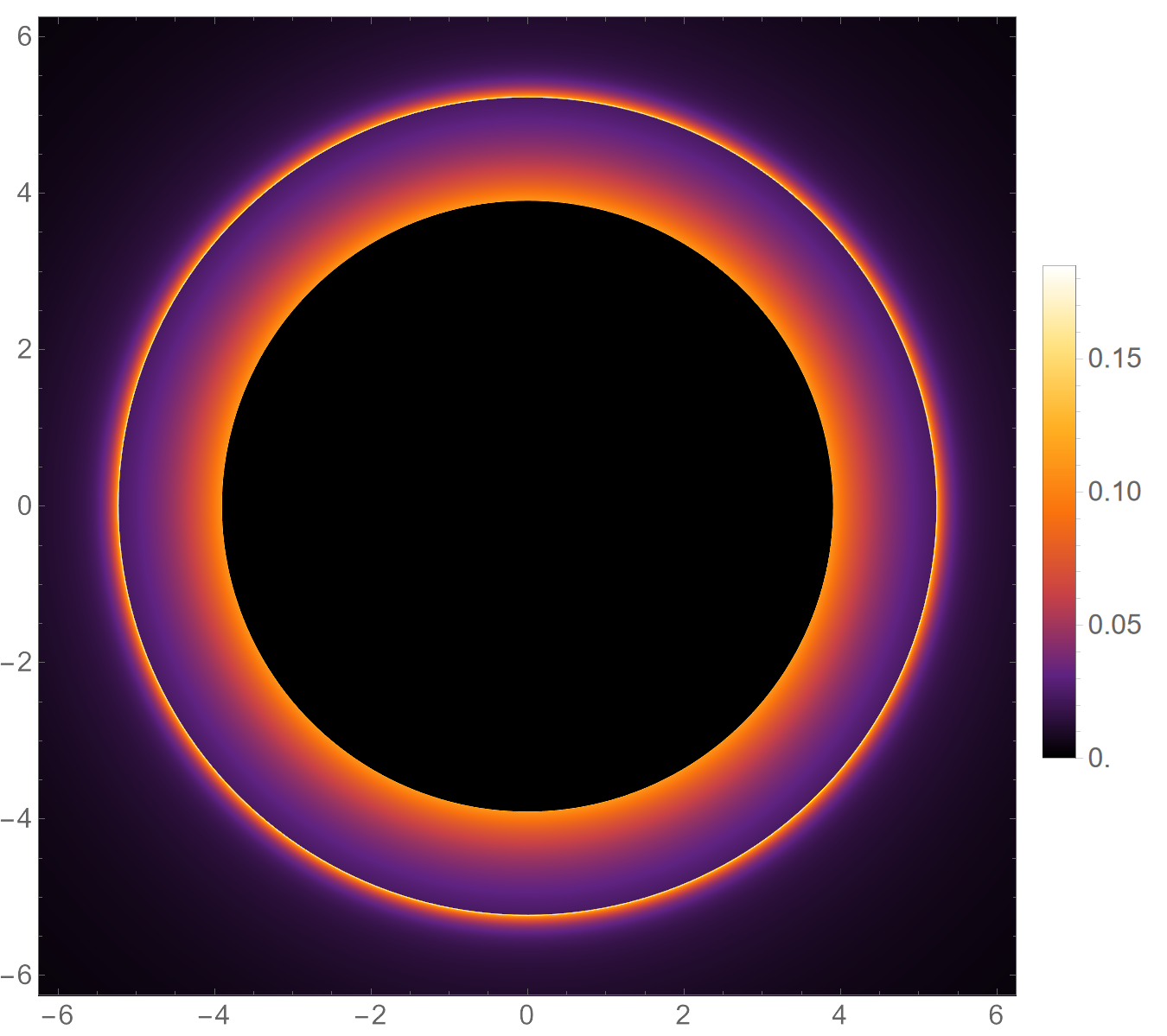}}

\caption{(a) The observed intensity $I_{obs}$. (b) The image of the regular black hole when the photons emitted in universe $B$.}
\label{rBHIobs}
\end{figure}

In Fig.\ref{rBHIobs}, we show the image of the regular black hole with the accretion disk and the observer both located at universe $B$.
The observed intensity is dominated by the direct emission.
This image is similar to the one of the Schwarzschild black hole \cite{Gralla:2019xty}.
They differ slightly in the radius of the photon ring and the size of the shadow.
However, it is hard to identify the location of the photon ring after blurring.
Thus it is difficult to distinguish image above and the Schwarzschild one because the metric outside the regular black hole is so similar to the Schwarzschild's.

Fortunately, the image whose accretion disk located at universe $A$ is significantly different from the Schwarzschild one.
And we will discuss it in next subsection.

\subsection{The image of the black hole formed by the photons emitted in universe $A$}
\begin{figure}[htb]
  \centering
  \includegraphics[width=5cm]{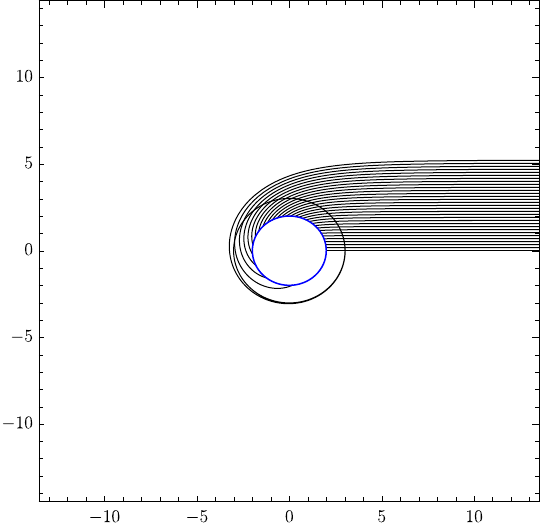}
  \caption{The trajectory of the photon in universe $B$}
  \label{geoin}
\end{figure}

Since the inner horizon is stable, the photons falling into the event horizon can cross the inner horizon.
Therefore, when the accretion disk is located at universe $A$, the photon with $b<b_c$ can also be received by the observer in universe $B$, as the black curve in Fig.\ref{geo}.
In this case, the trajectory of the photon outside the white hole in universe $B$ is shown in Fig.\ref{geoin}.
The inner circle and the outer circle are the event horizon and the photon sphere of the white hole in universe $B$, respectively.

\begin{figure}[htb]
  \centering
  \includegraphics[width=5cm]{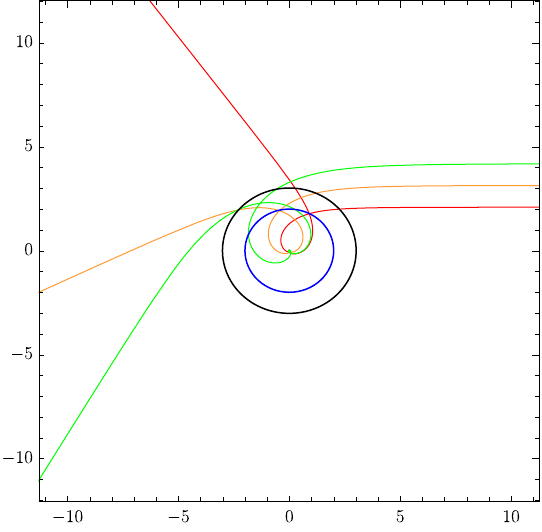}
  \caption{The trajectory of the photons from universe $A$ to universe $B$. The black circle and the blue circle are the photon sphere and the event horizon of the black (white) hole.}
  \label{geoAB}
\end{figure}

If we overlap the regular black hole in universe $A$ and the white hole in universe $B$ together, then the total trajectory of photons are shown as Fig.\ref{geoAB}.

\begin{figure}[htbp]
\centering
\subfigure[]{\includegraphics[width=4cm]{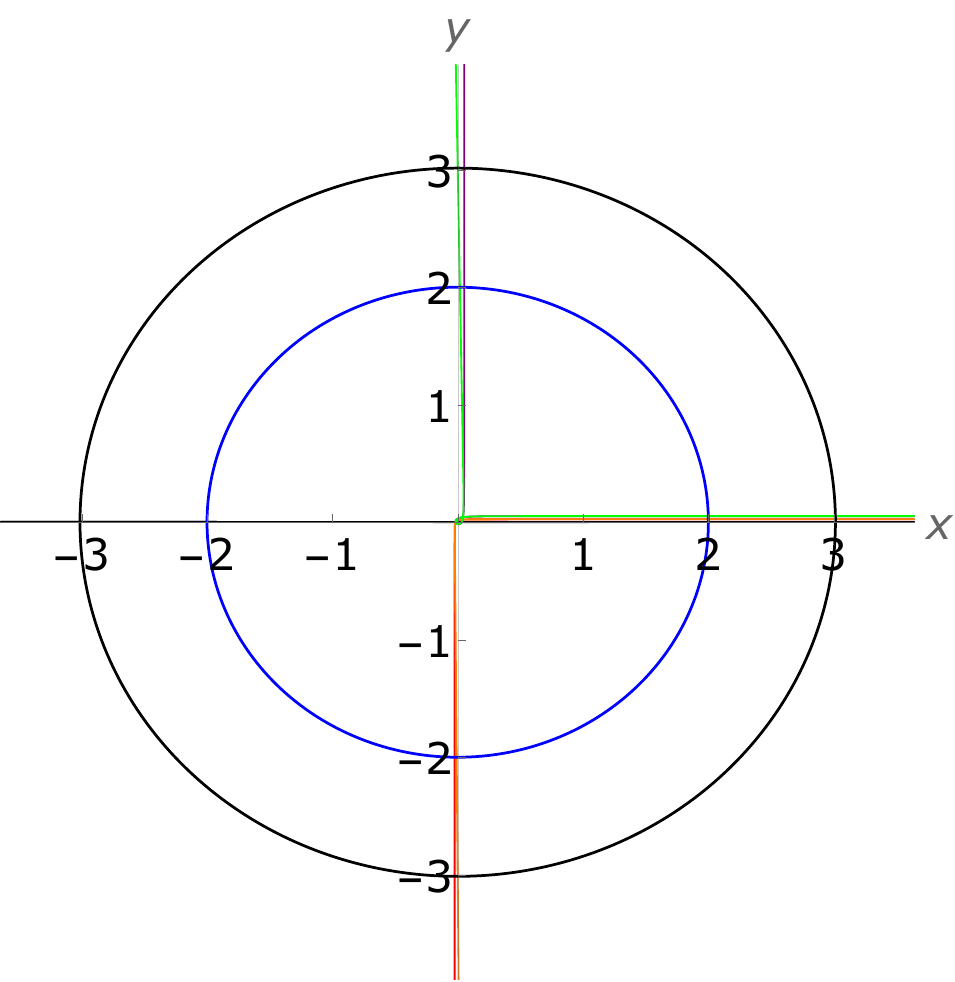}}
\hspace{0.5cm}
\subfigure[]{\includegraphics[width=6cm]{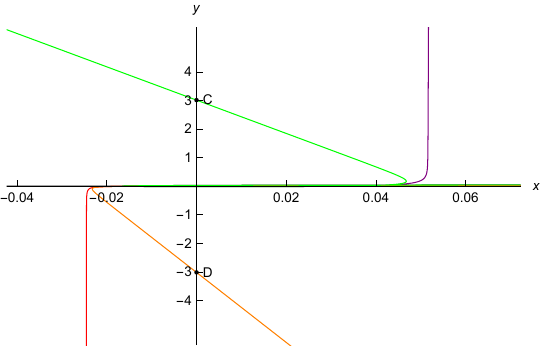}}
\hspace{0.5cm}
\subfigure[]{\includegraphics[width=6cm]{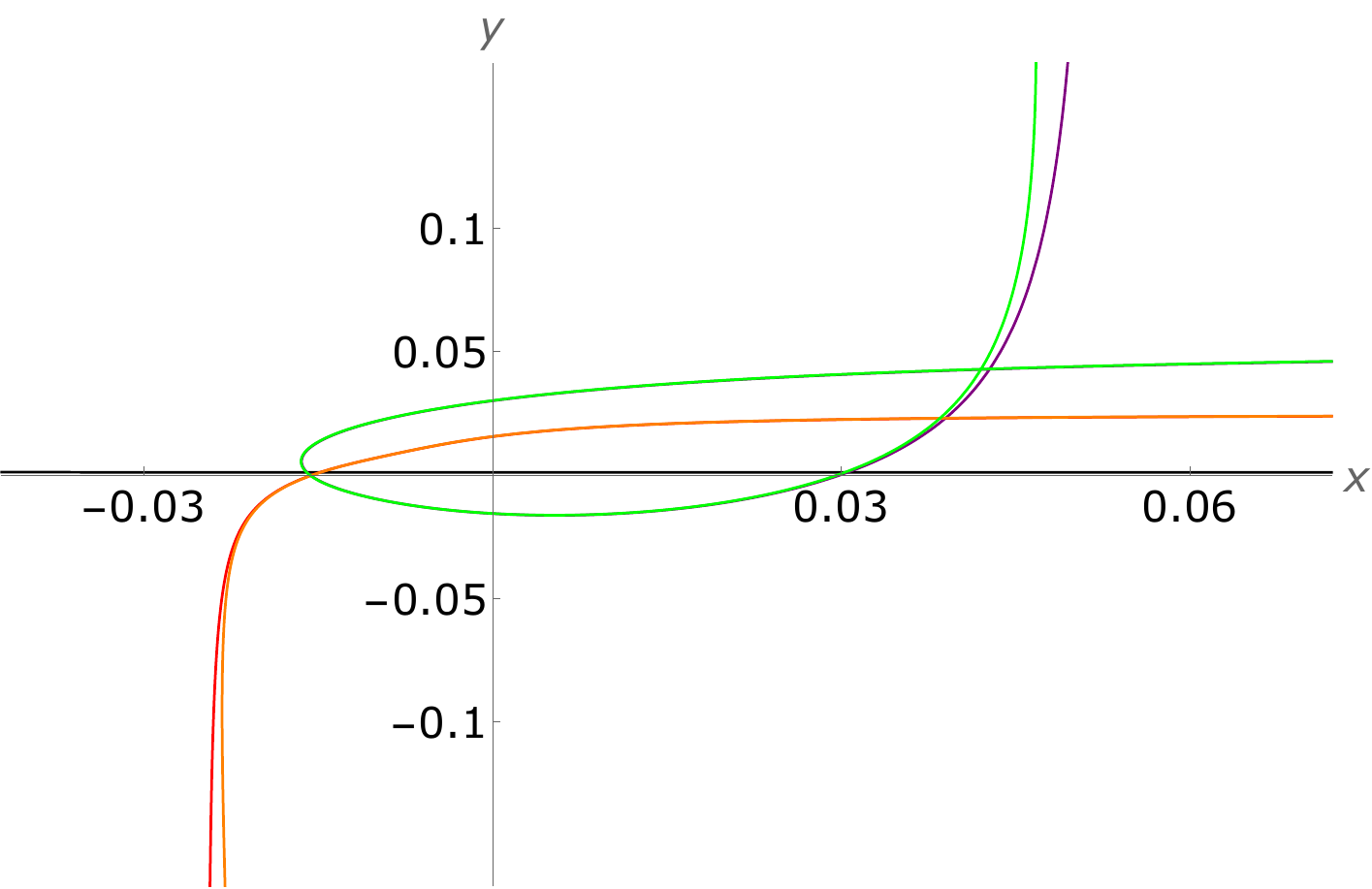}}
\caption{The trajectory of the photons with some critical $n$.
All the ray end at $x\rightarrow+\infty$.
Picture (b) shows the stretching of the $x$-axis in picture (a).
And picture (c) shows the stretching of the $y$-axis in picture (b).
The black circle is the photon sphere and the blue circle is the event horizon.
The red curve and purple curve correspond to $n=3/4$ and $5/4$.
The orange curve intersect with the accretion disk and the photon sphere at the same point $D$.
And the green curve intersect with them at $C$.}
\label{geoAB1}
\end{figure}

Especially, we draw the ray with $n=3/4$, and $n=5/4$ in Fig.\ref{geoAB1}.
When $b$ tends to zero, the ray nearly overlaps with the $x$-axis.
As we increase $b$, the ray deflects by a small angle.
The change of the azimuthal angle $\phi$ of a ray begins and ends at infinity is
\begin{eqnarray}
\label{angle}
\Delta \phi=2\int_{r_{\text{min}}}^{\infty}\frac{\mathrm{d}r}{r^2\sqrt{b^{-2}-V(r)}}\,.
\end{eqnarray}
Therefore, as $b$ increases, the azimuthal angle increase, too.
When $b=0.02454$, the azimuthal angle $\phi$ of the starting point of the ray is $(3/2)\pi$, or $n=3/4$, as the red curve.
If we increase $b$ slightly, then the ray will have a larger deflection angle, and intersects with the accretion disk.
This kind of ray can be received by the screen, until $b=0.02461$ as the orange curve.
Therefore, a bright ring occurs on the screen.
However, as $b$ continues to increase, the ray will not intersect with the accretion disk.
Until the ray intersects with the accretion disk again at the top half of the $y$-axis with $n\gtrsim5/4, b\gtrsim0.05158$ as the purple curve, and the second bright ring occurs.
This second ring end at $b=0.05174$ as the green curve, which intersect with the accretion disk at point $C$.
In the same manner, there are many rings appear on the screen with $b<b_c$.
As the value of $r$ approaches zero, $V(r)$ tends to $1/r^2$.
Consequently, the trajectory of the light ray we mentioned above behaves similar to that of Cotes's spirals \cite{Whittaker:1988}.
Although there being only a single turning point, the light ray has the capability to rotate around the center multiple times.

\begin{figure}[htbp]
\centering
\subfigure[]{\includegraphics[width=8cm]{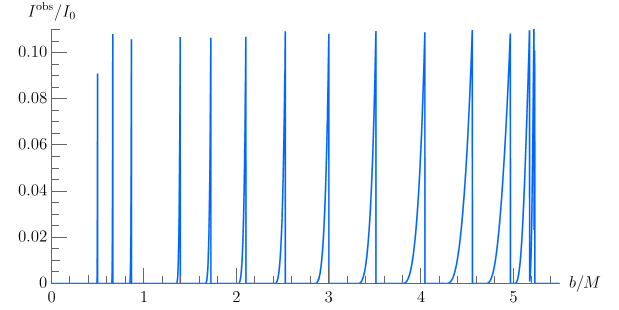}}
\subfigure[]{\includegraphics[width=5cm]{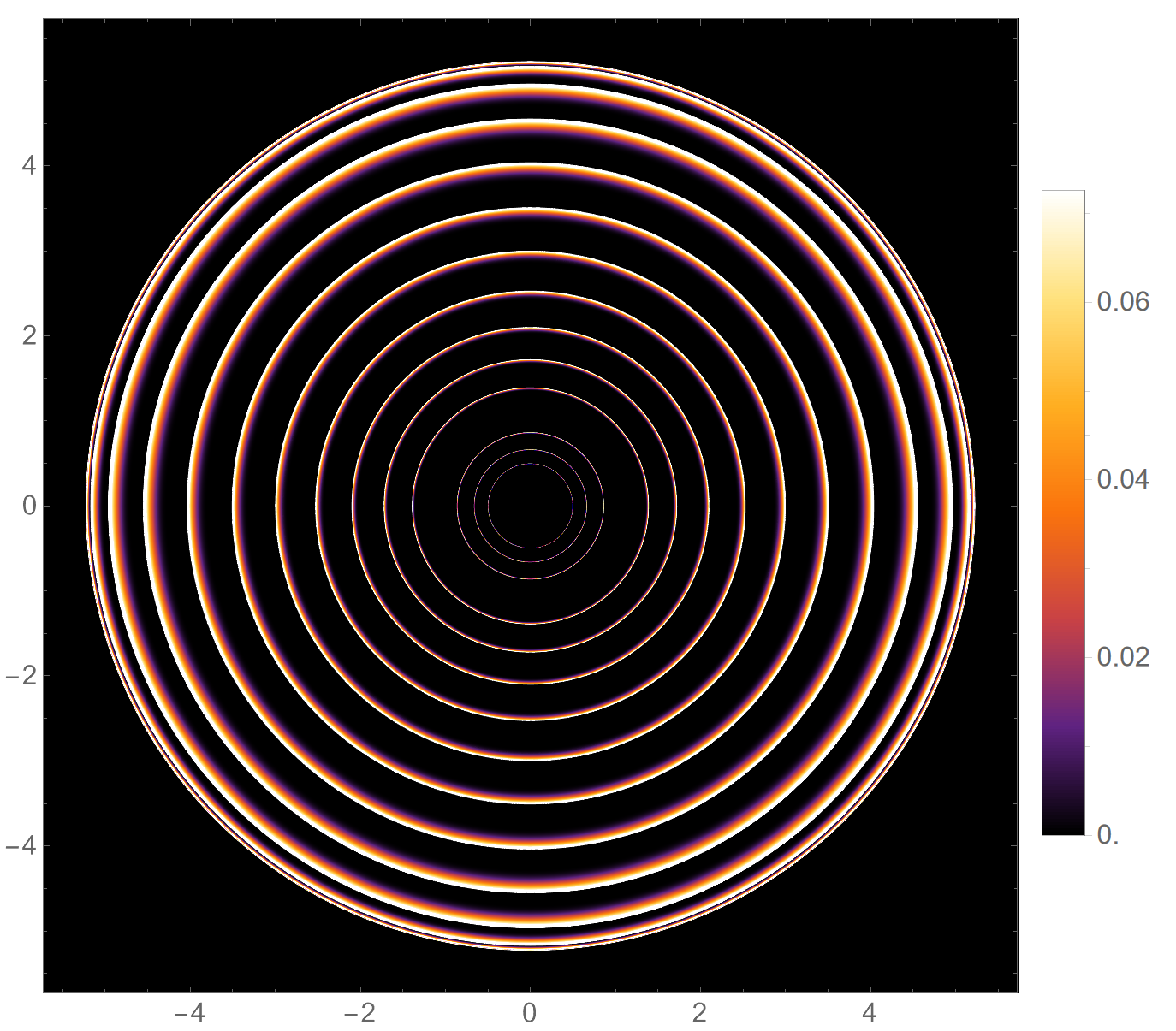}}
\caption{The observed intensity (a) and the appearance (b) when the accretion disk is located at universe $A$ while the observer is in universe $B$.}
\label{50}
\end{figure}

The observed intensity and the image are depicted in Fig.\ref{50}.
As we can see, the multi-ring structure occurs for the regular black hole.
Unlike the image of the Schwarzschild black hole, where there is only one or two bright rings lying in the region $b\gtrsim b_c$ when the accretion disk located outside the photon sphere, in our case, the bright rings with $n>5/4$ and $b<b_c$ are distinctly separated, allowing us to distinguish them.
When $b$ tends to $b_c$, the bright rings become more and more concentrated and indistinguishable.
This multi-ring structure also appears in some other massive bodies without horizon, such as compact object and wormhole \cite{Olmo:2021piq,Guerrero:2022qkh,Guerrero:2022msp,Olmo:2023lil}.
In fact, when the effective potential $V(r)$ of the spacetime have a barrier, the photon sphere occurs and there is a bright ring in the image.
Furthermore, if there is an another higher barrier inside the outer one, or $V$ is diverging at somewhere inside the photon sphere, then the multi-ring structure appears.
In \cite{Zhang:2023okw}, the image of the quantum-corrected black hole is investigated, and it has the similar multi-ring structure.
However, the inner horizon of the quantum-corrected black hole is unstable and will turn into a null singularity \cite{Cao:2023aco}.
Therefore the photons are unable to pass through the inner horizon and escape from a white hole.
Due to the instability of its inner horizon and the slightness of the quantum correction, it may be challenging to detect this novel appearance of the quantum-corrected black hole.
Nevertheless, the inner horizon of the regular black hole we are studying is stable which give rise to the multi-ring structure.
It also means that the multi-ring structure can occur in the appearance of the massive object with horizon.

\begin{figure}[htbp]
\centering
\subfigure[]{\includegraphics[width=8cm]{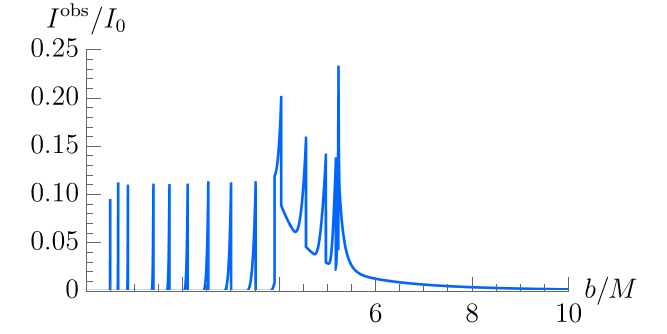}}
\subfigure[]{\includegraphics[width=5cm]{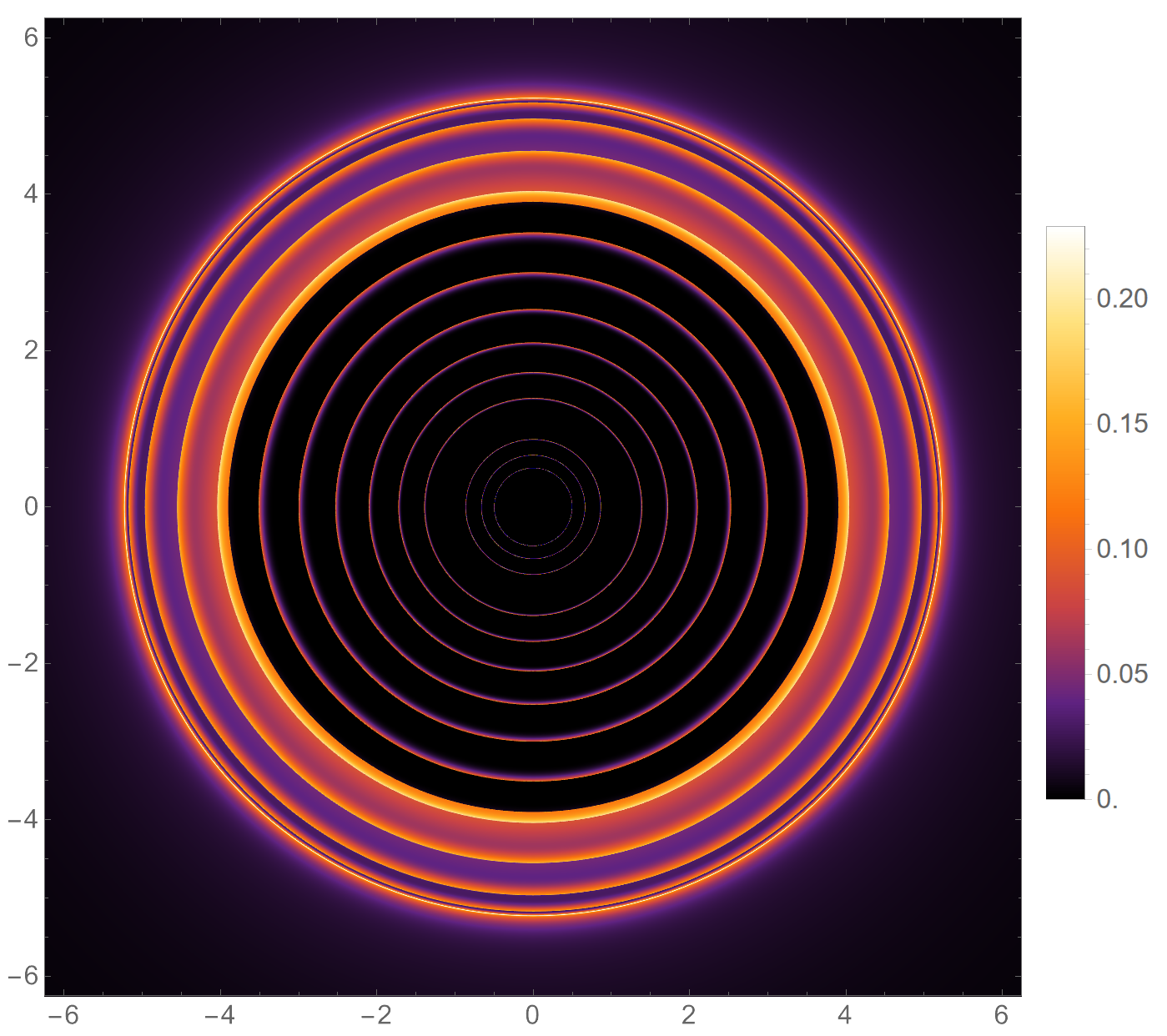}}
\caption{The observed intensity (a) and the appearance (b) when the accretion disks are located both the two universes. }
\label{50tot}
\end{figure}

In some special cases, there is an accretion disk in universe $A$ and another one in universe $B$ at the same time.
The intensity of them are both illustrated in Fig.\ref{Iem}.
Then the image of the regular black (white) hole is the overlay of Fig.\ref{rBHIobs} and Fig.\ref{50}, and it is shown as Fig.\ref{50tot}.
The total luminosity is largely dominated by the direct emission.
It has many rings, which possesses significant difference with Schwarzschild black hole.
Especially, some rings appear distinctly in the shadow region.
We can distinguish the regular black hole from the Schwarzschild black hole by this multi-ring structure.
It is worth noting that, the radius of the rings with $b<b_c$ only depends on the geometry of the black hole, not on the structure of the accretion disks.
This provides us with an effective method for quantifying the physical properties of a regular black hole.

\section{The appearance of the regular black hole with different parameters}\label{S4}
The regular black hole we are studying has four parameters, i.e., $r_+, r_-, M $, and $b_2$. They have a significant impact on image of the regular black hole.
To compare with the Schwarzschild black hole, we fix $M=1$ and $r_+=2$.

\begin{figure}[htbp]
\centering
\subfigure[]{\includegraphics[width=4.5cm]{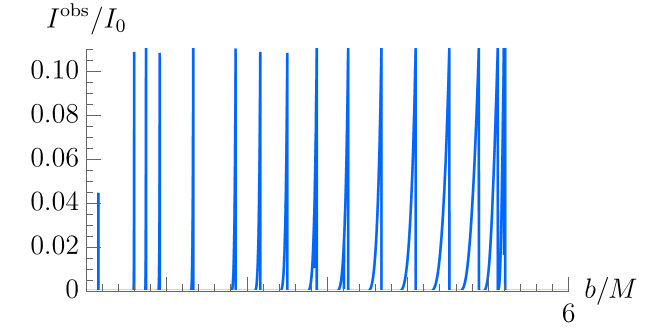}}
\subfigure[]{\includegraphics[width=3cm]{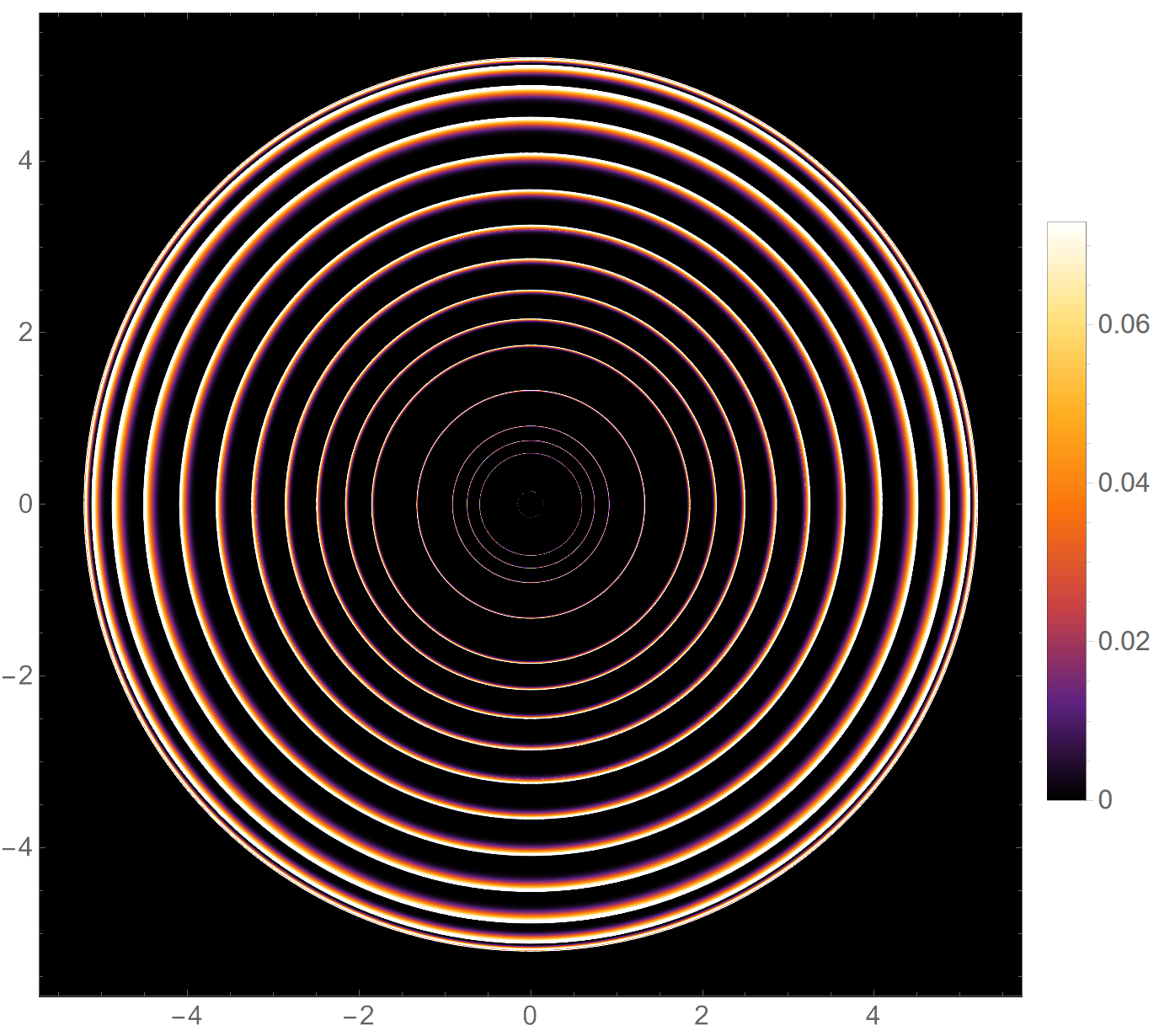}}
\subfigure[]{\includegraphics[width=4.5cm]{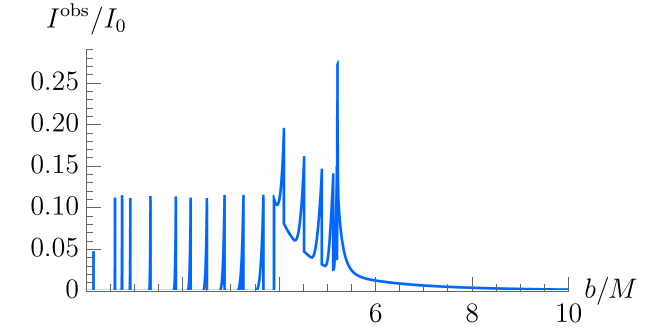}}
\subfigure[]{\includegraphics[width=3cm]{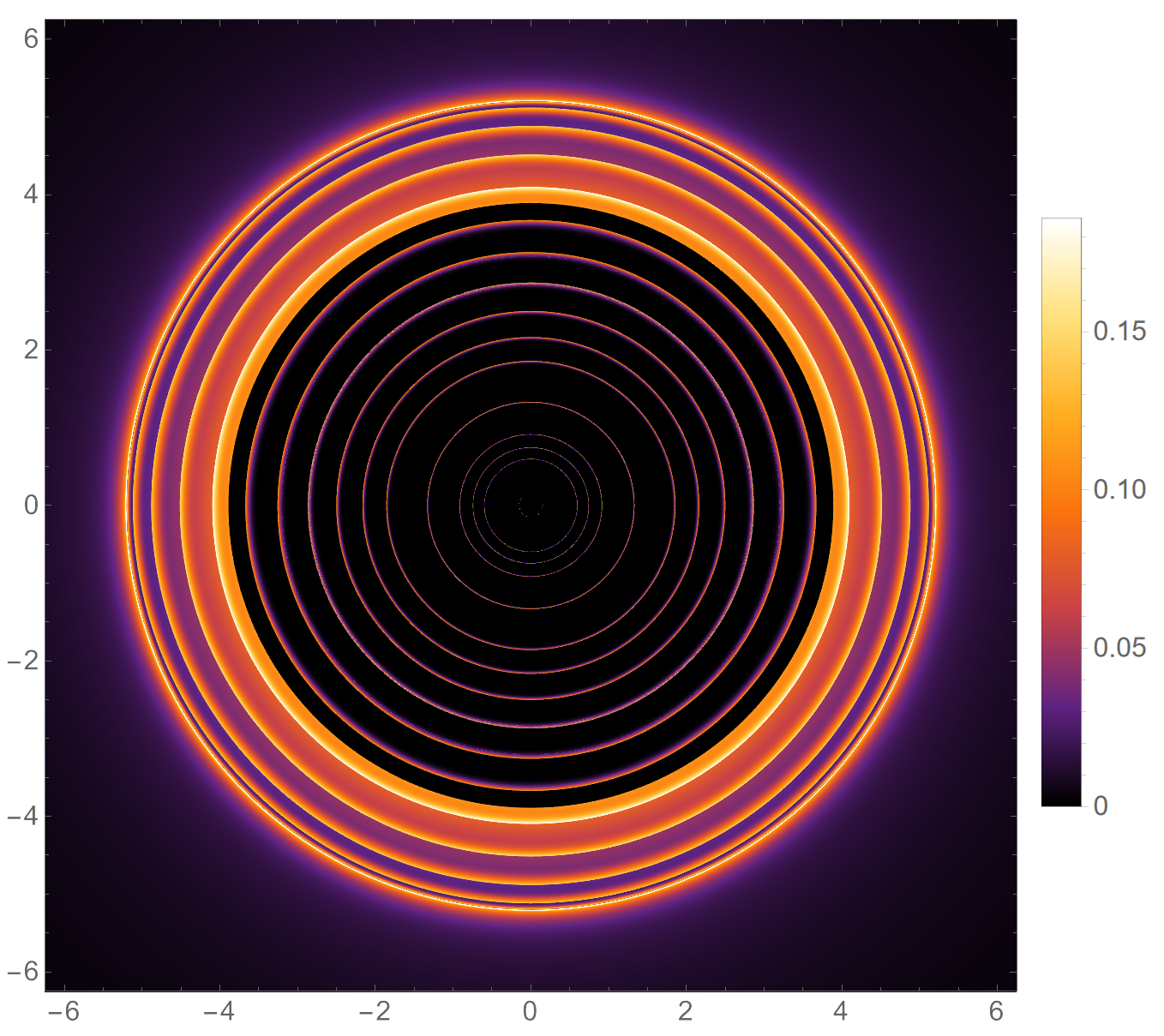}}

\subfigure[]{\includegraphics[width=4.5cm]{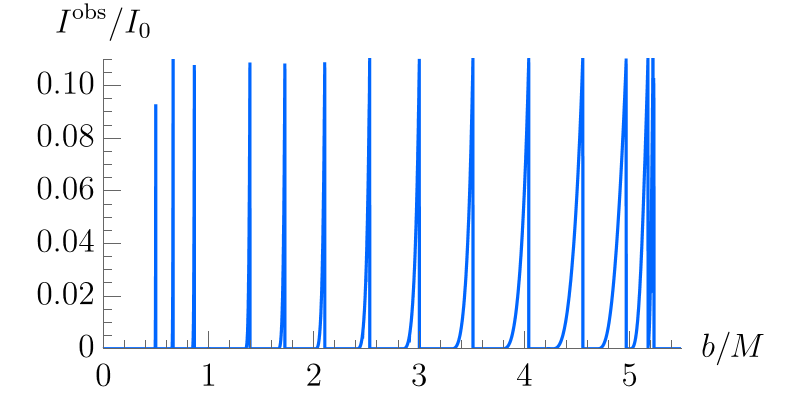}}
\subfigure[]{\includegraphics[width=3cm]{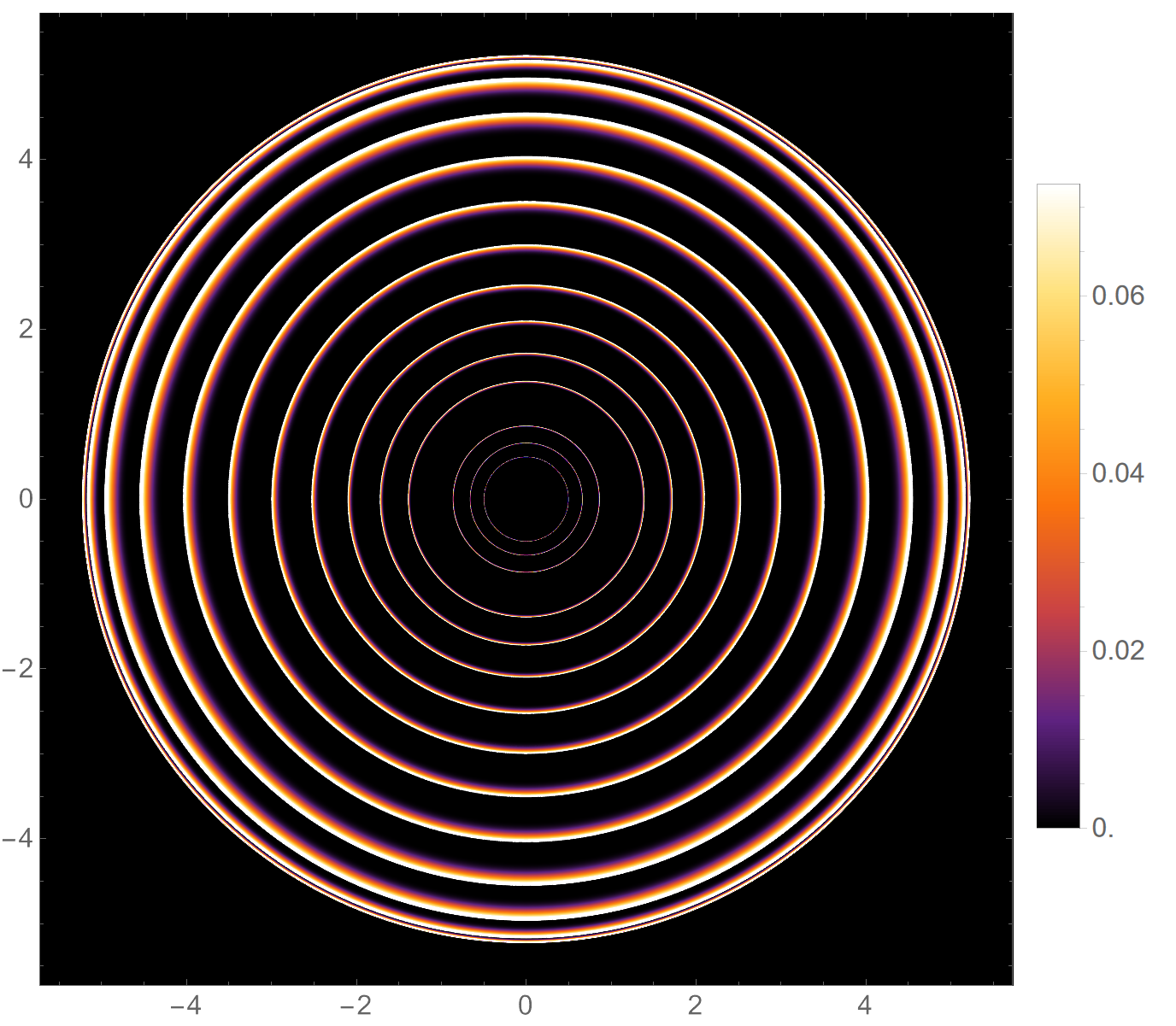}}
\subfigure[]{\includegraphics[width=4.5cm]{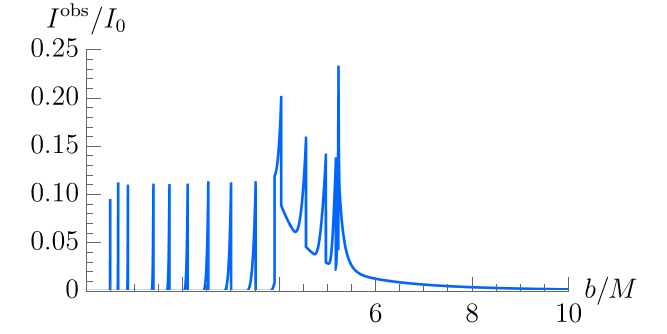}}
\subfigure[]{\includegraphics[width=3cm]{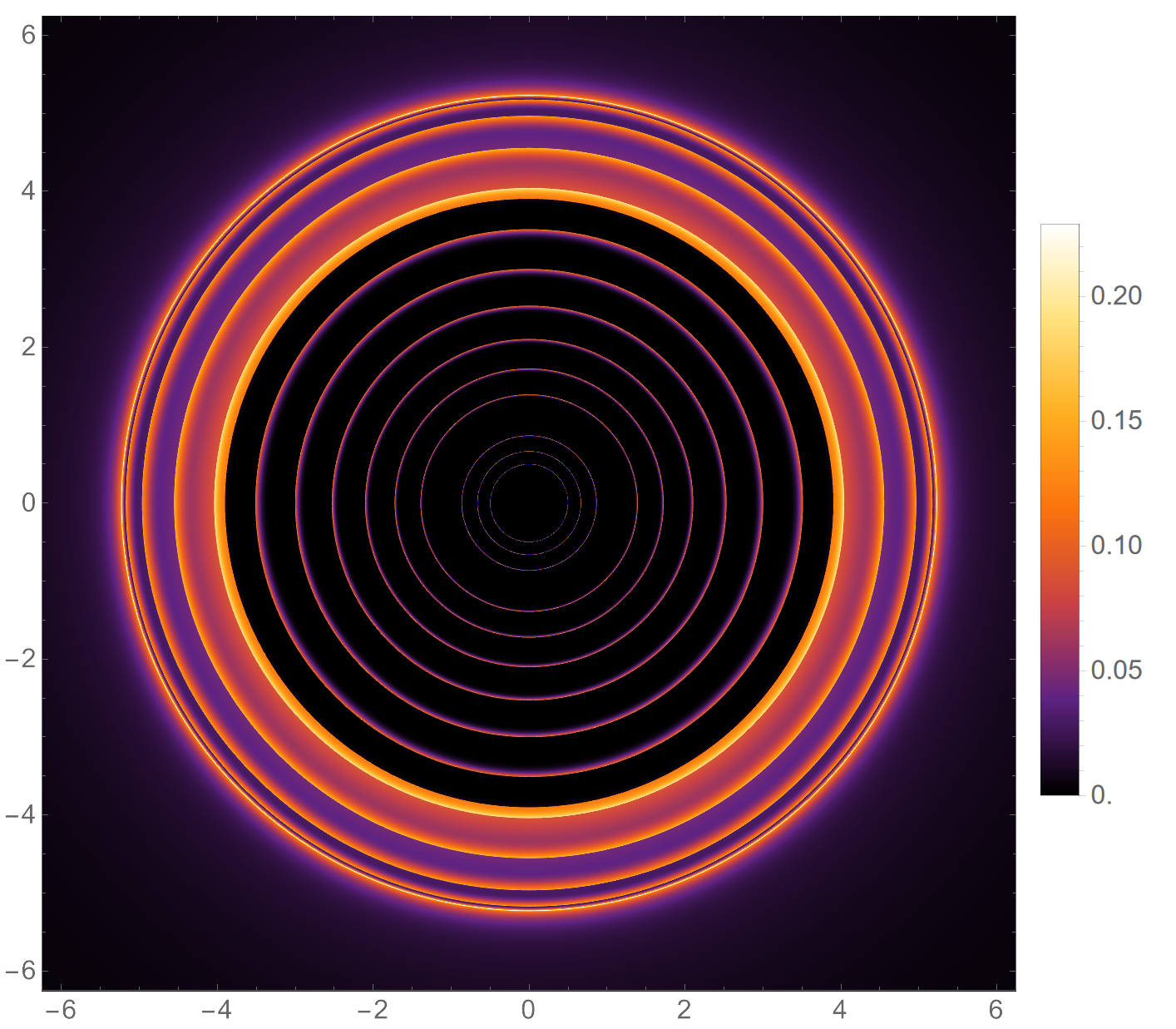}}

\subfigure[]{\includegraphics[width=4.5cm]{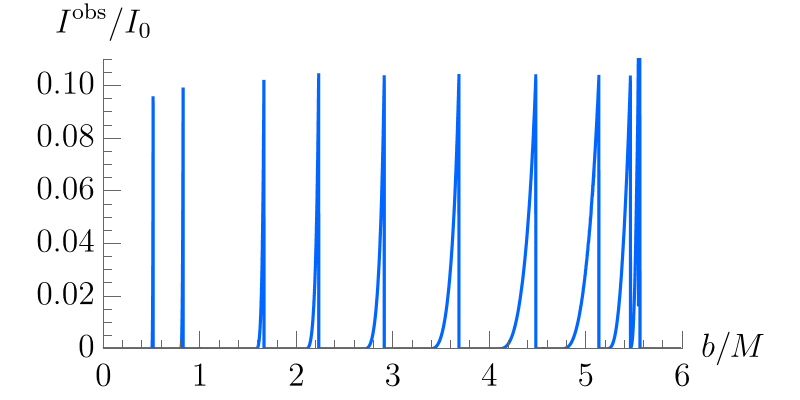}}
\subfigure[]{\includegraphics[width=3cm]{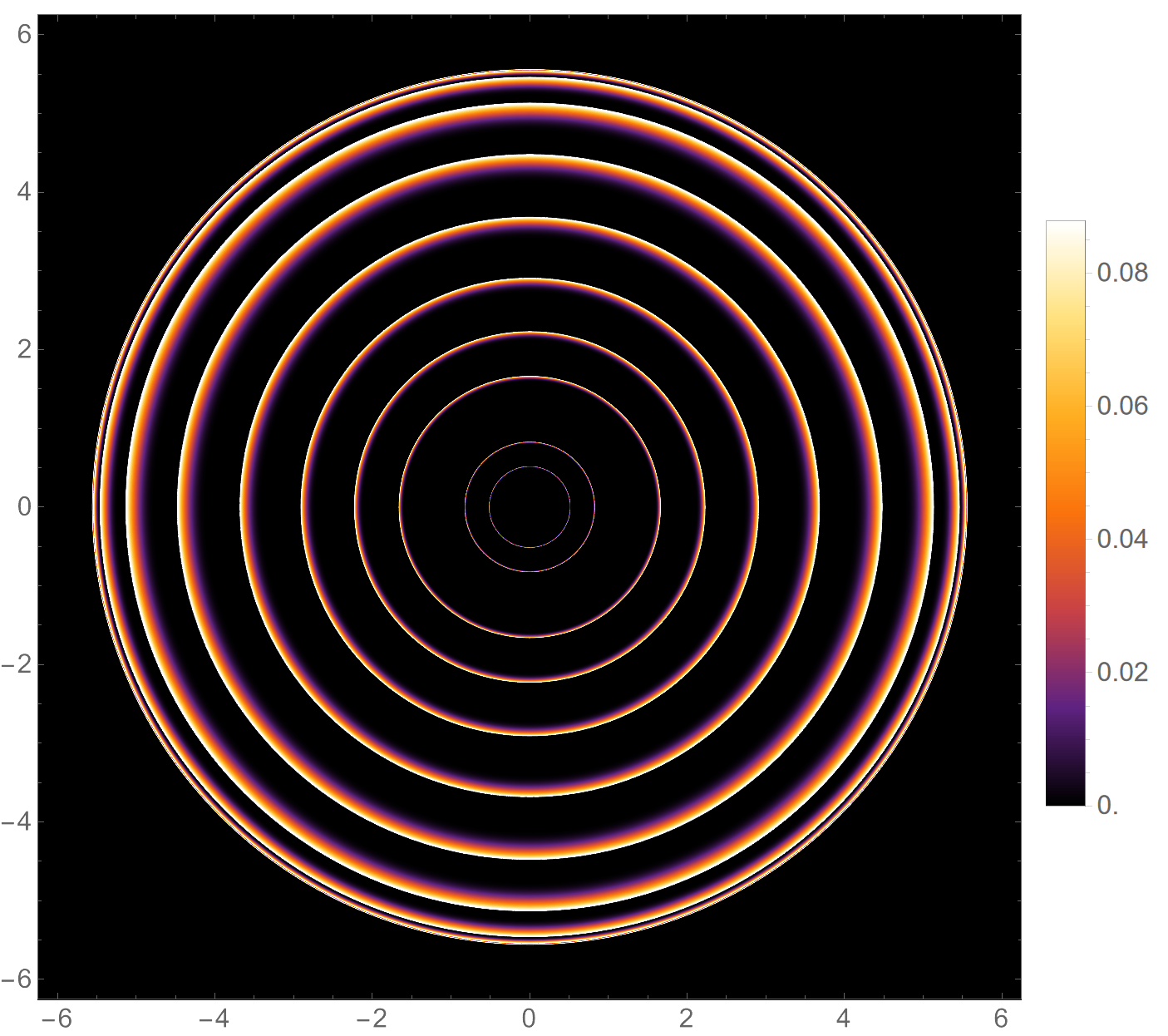}}
\subfigure[]{\includegraphics[width=4.5cm]{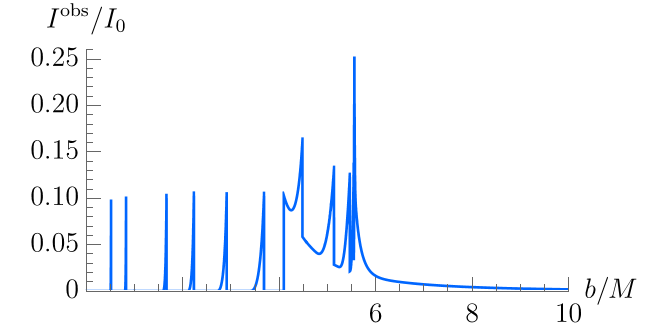}}
\subfigure[]{\includegraphics[width=3cm]{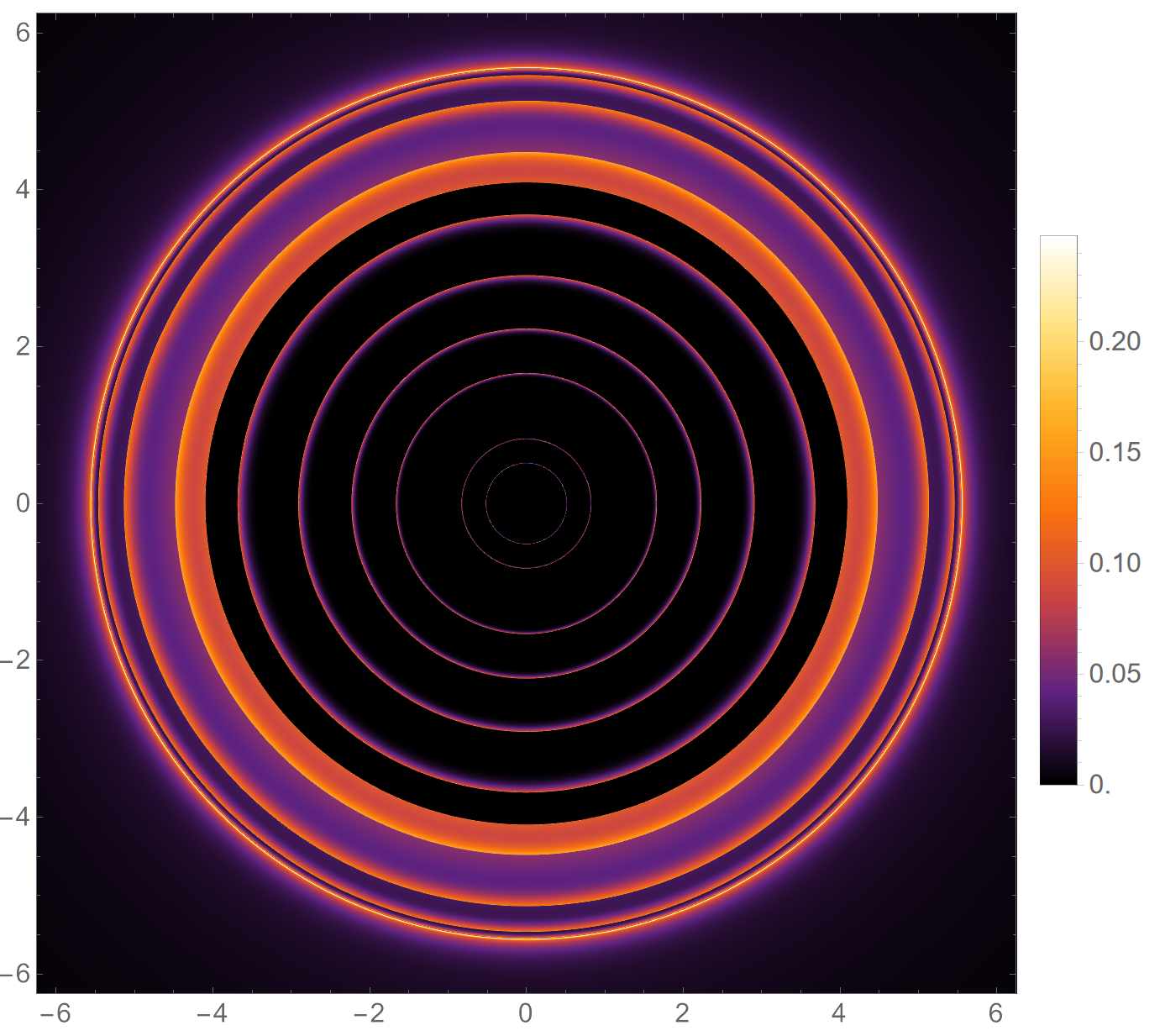}}

\caption{The observed intensity and the appearance for the regular black hole with $r_-=0.008$ (top), $r_-=0.02$ (middle) and $r_-=0.2$ (bottom).
The other three parameters of the regular black hole are $r_+=2, M=1, b_2=0$.
The first column is the observed intensity when the accretion disk is only located in universe $A$ while the observer is in universe $B$.
And the second column is the corresponding image.
The third column is the observed intensity when the accretion disks are located both in universe $A$ and $B$.
Its image is shown as the fourth column.}
\label{rm}
\end{figure}

In Fig.\ref{rm}, we have fixed $b_2=0$ to examine the effect of $r_-$ on the observed intensity and the appearance of black hole.
As we can see, if we only consider the photons emitted in the universe $A$ (case $A$), i.e. the left column, the number of rings decreases when $r_-$ increases.
While when the photons only emit in universe $B$ (case $B$), the observed intensity and the appearance are similar to that of the Schwarzschild black hole as we expected since the Schwarzschild black hole is the limiting case for the regular black hole in some sense.
The only differences are the specific locations of peaks and corresponding intensities.

When these parameters take some particular values, the Schwarzschild black hole and regular black hole in case $B$ are at least observationally indistinguishable.
As a result, when we consider both cases, i.e., the accretion disk located both universe $A$ and $B$ (case $AB$), what we get is the intensity in case $A$ superimposed on an intensity that is almost unchanging.
So the second column and the fourth column have the same tendencies.
As $r_-$ increases, the number of rings decreases.

\begin{figure}[htbp]
\centering

\subfigure[]{\includegraphics[width=4.5cm]{r02WHt.png}}
\subfigure[]{\includegraphics[width=3cm]{r02WH.png}}
\subfigure[]{\includegraphics[width=4.5cm]{r02BHWHt.png}}
\subfigure[]{\includegraphics[width=3cm]{r02BHWH.png}}

\subfigure[]{\includegraphics[width=4.5cm]{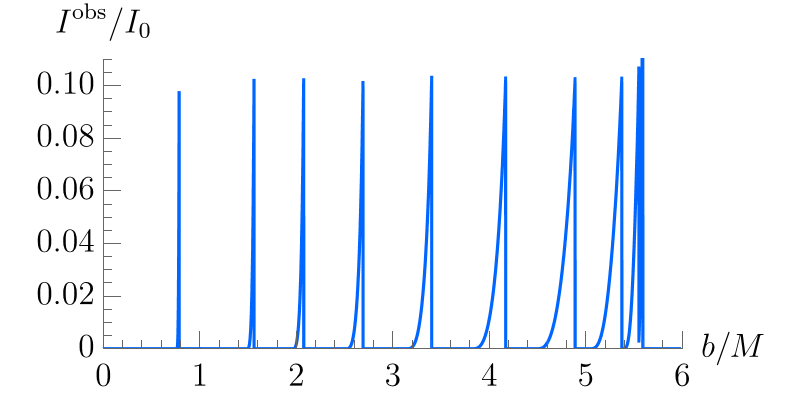}}
\subfigure[]{\includegraphics[width=3cm]{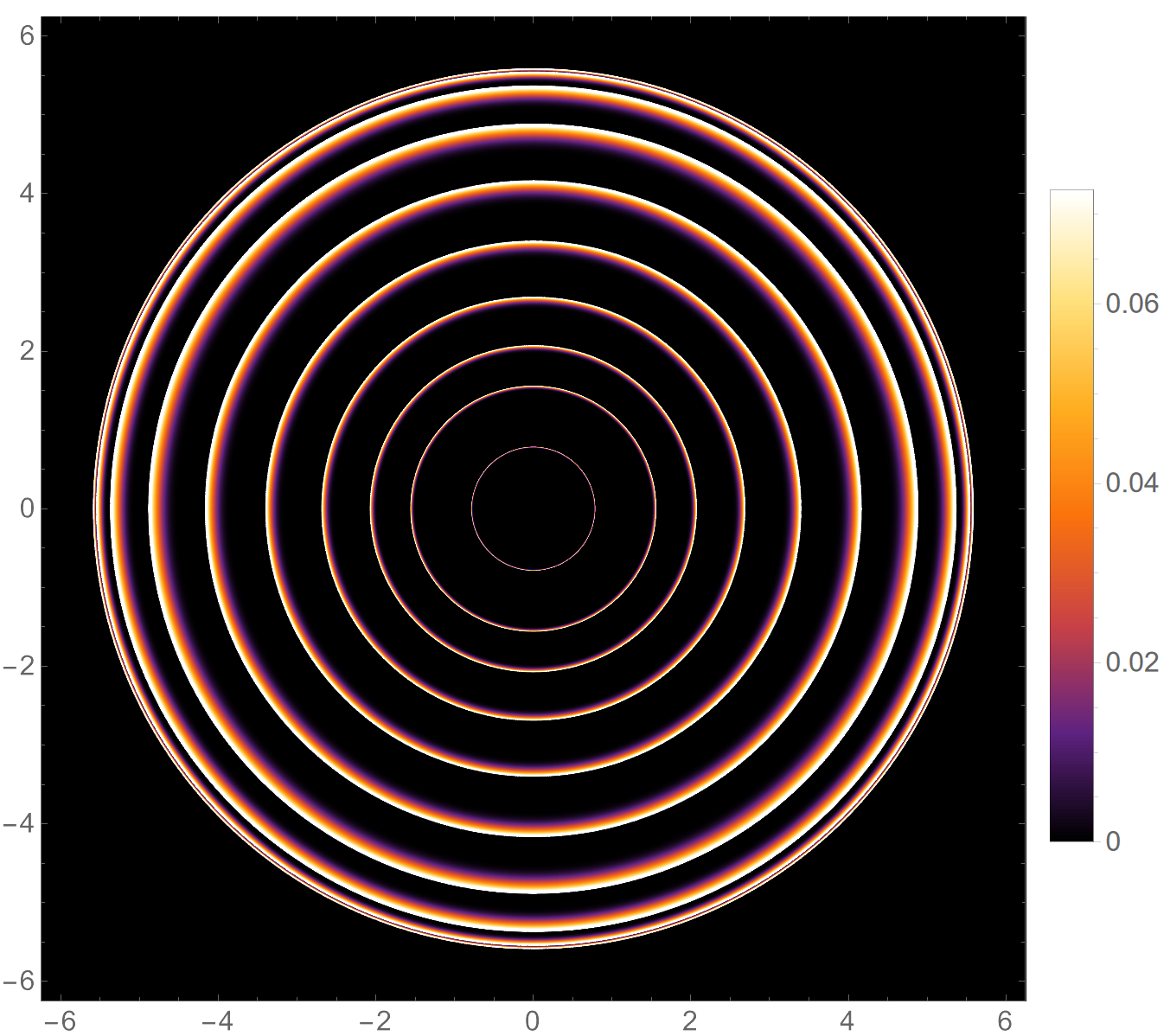}}
\subfigure[]{\includegraphics[width=4.5cm]{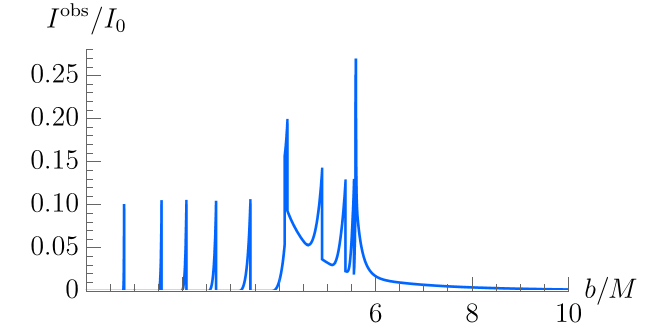}}
\subfigure[]{\includegraphics[width=3cm]{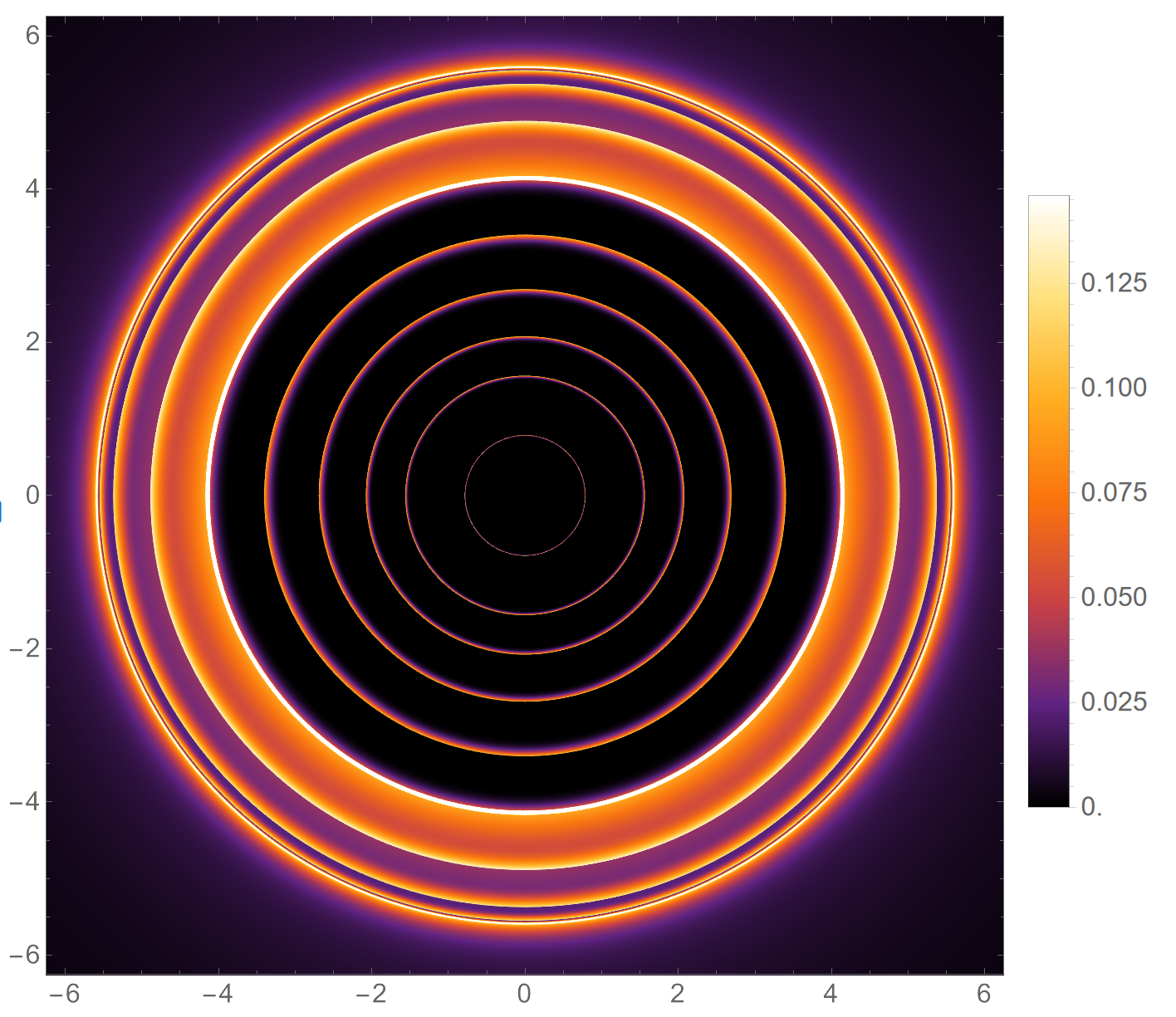}}

\subfigure[]{\includegraphics[width=4.5cm]{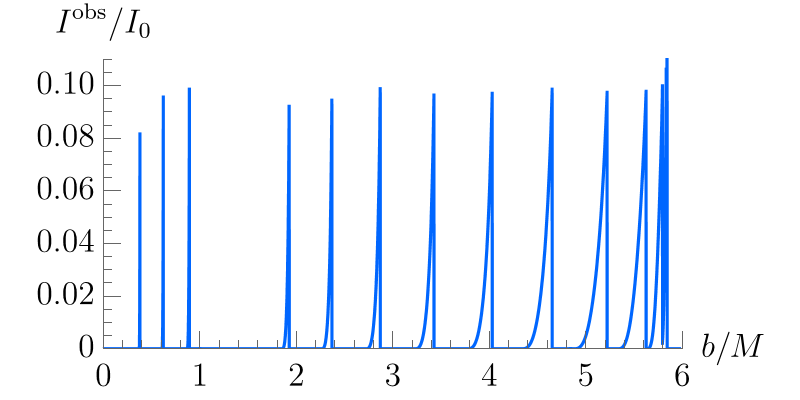}}
\subfigure[]{\includegraphics[width=3cm]{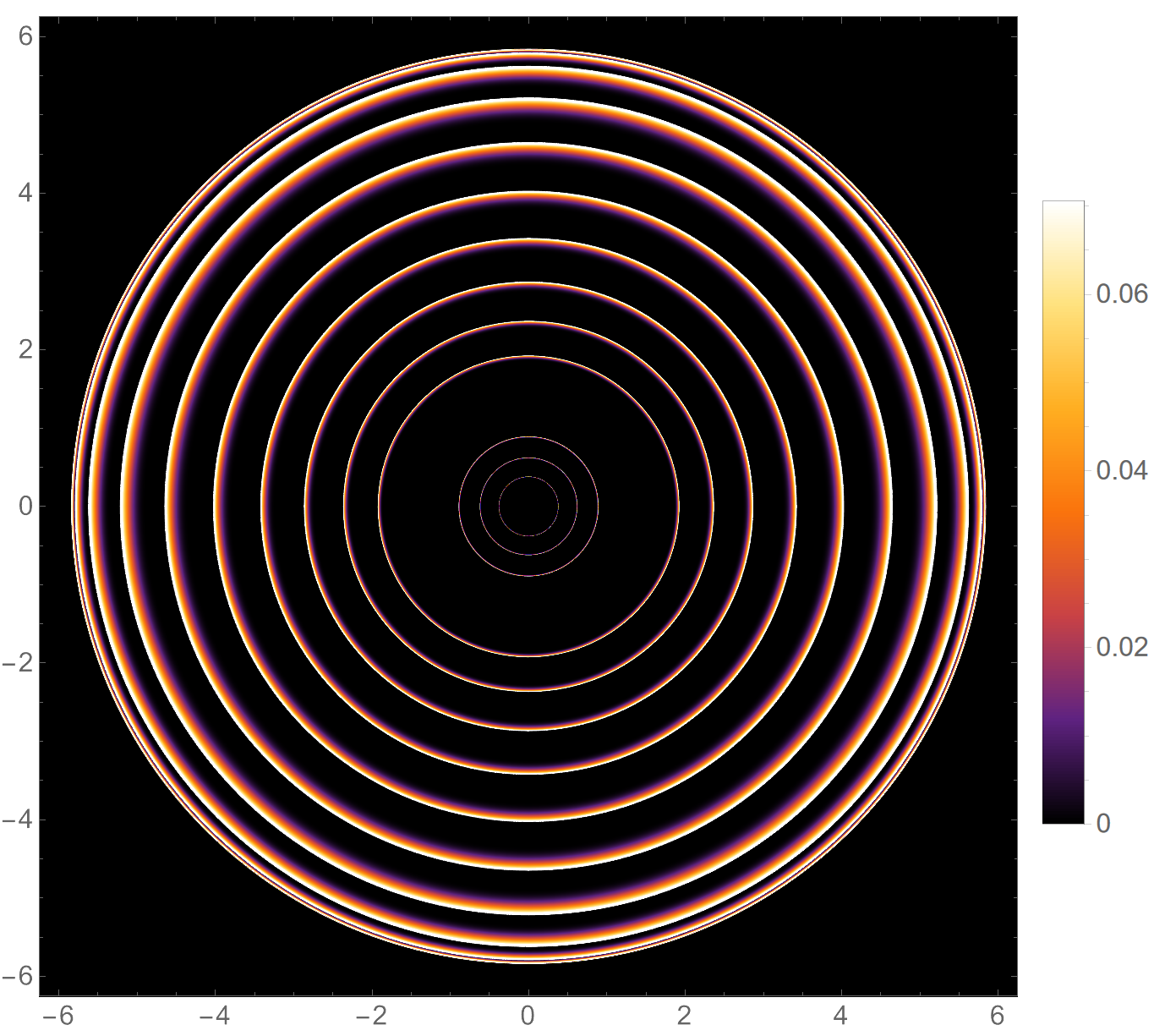}}
\subfigure[]{\includegraphics[width=4.5cm]{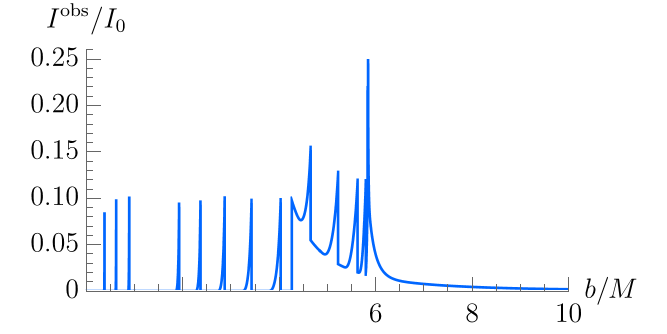}}
\subfigure[]{\includegraphics[width=3cm]{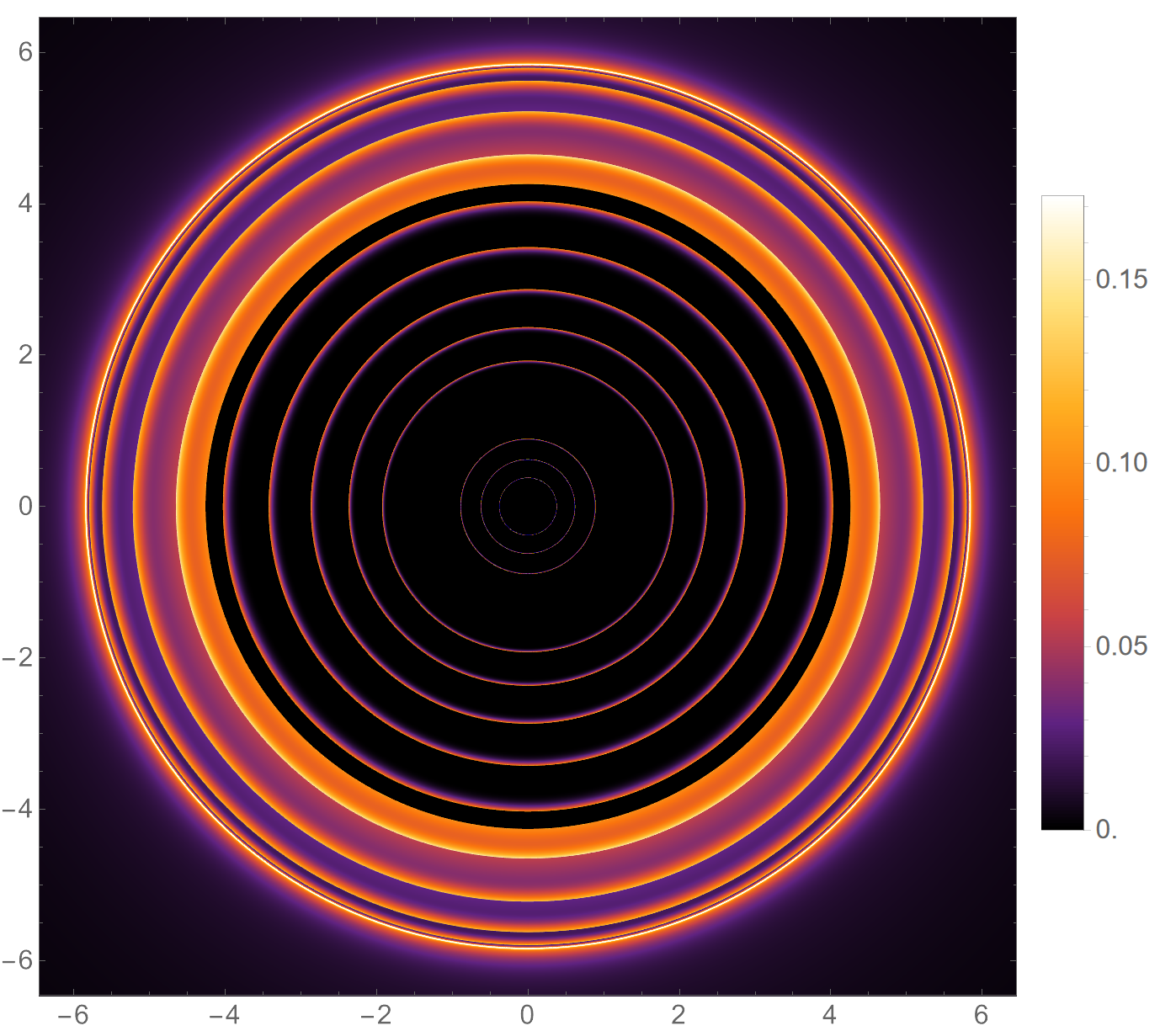}}

\subfigure[]{\includegraphics[width=4.5cm]{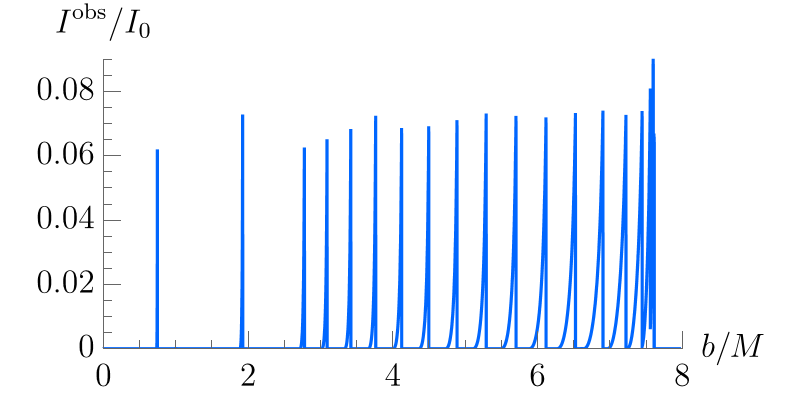}}
\subfigure[]{\includegraphics[width=3cm]{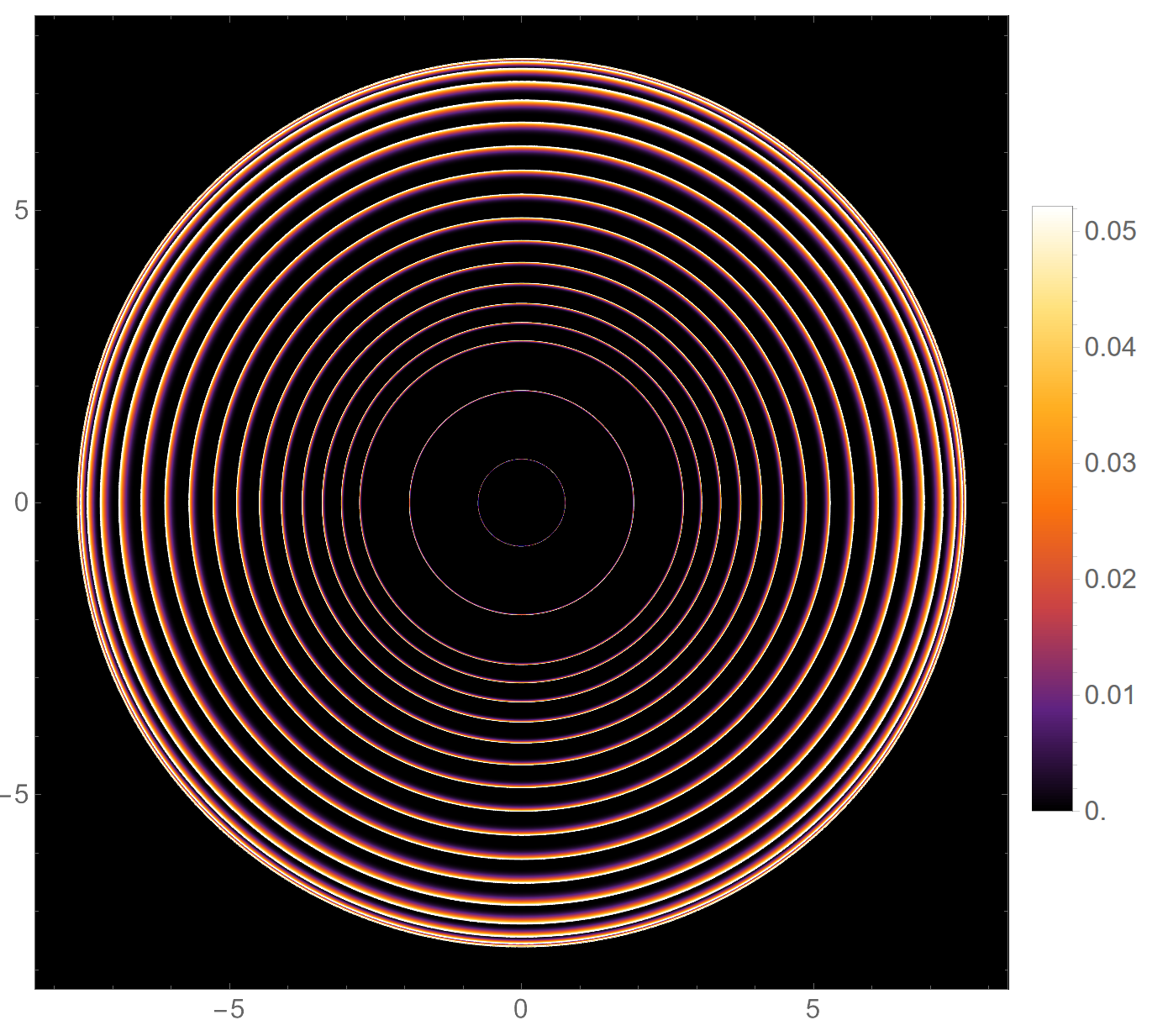}}
\subfigure[]{\includegraphics[width=4.5cm]{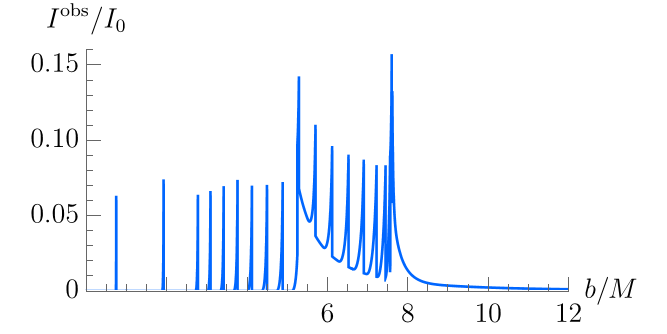}}
\subfigure[]{\includegraphics[width=3cm]{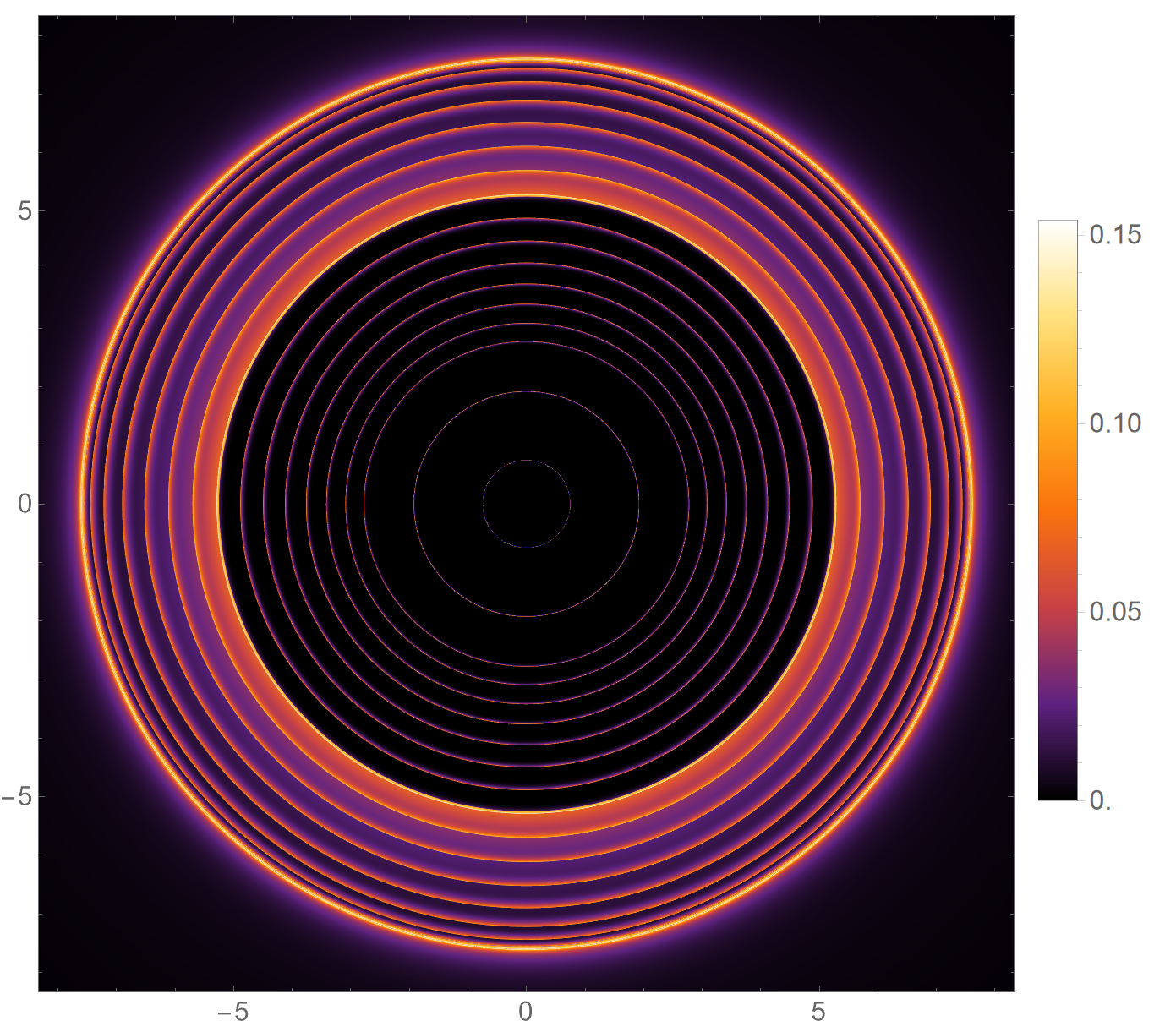}}

\subfigure[]{\includegraphics[width=4.5cm]{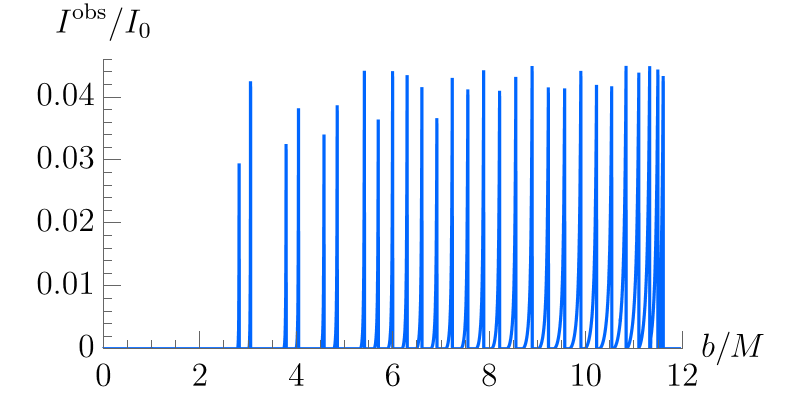}}
\subfigure[]{\includegraphics[width=3cm]{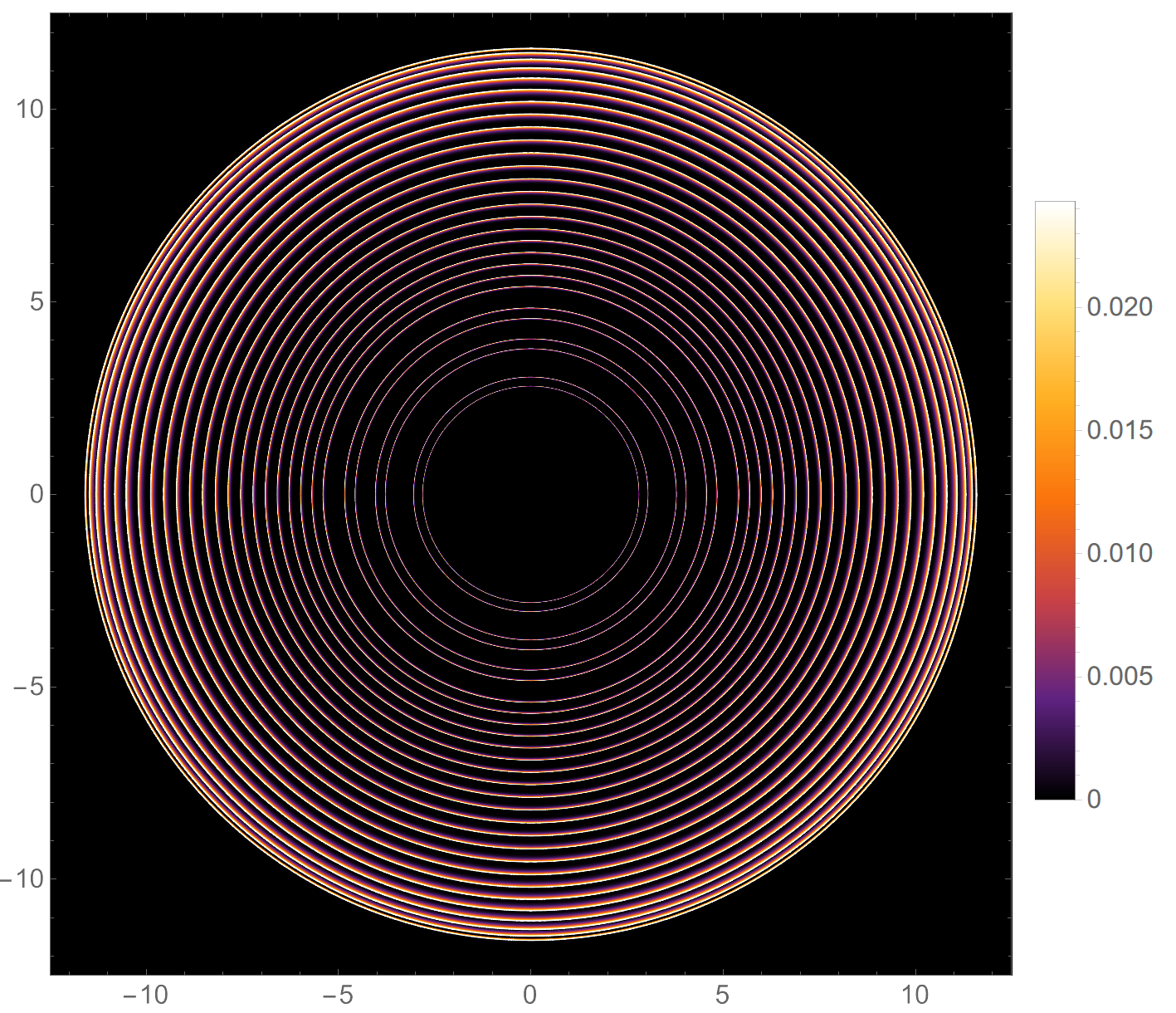}}
\subfigure[]{\includegraphics[width=4.5cm]{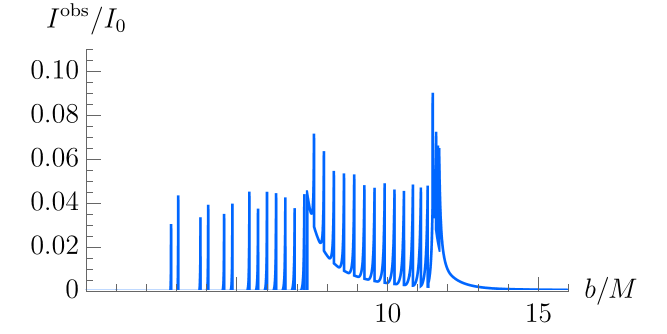}}
\subfigure[]{\includegraphics[width=3cm]{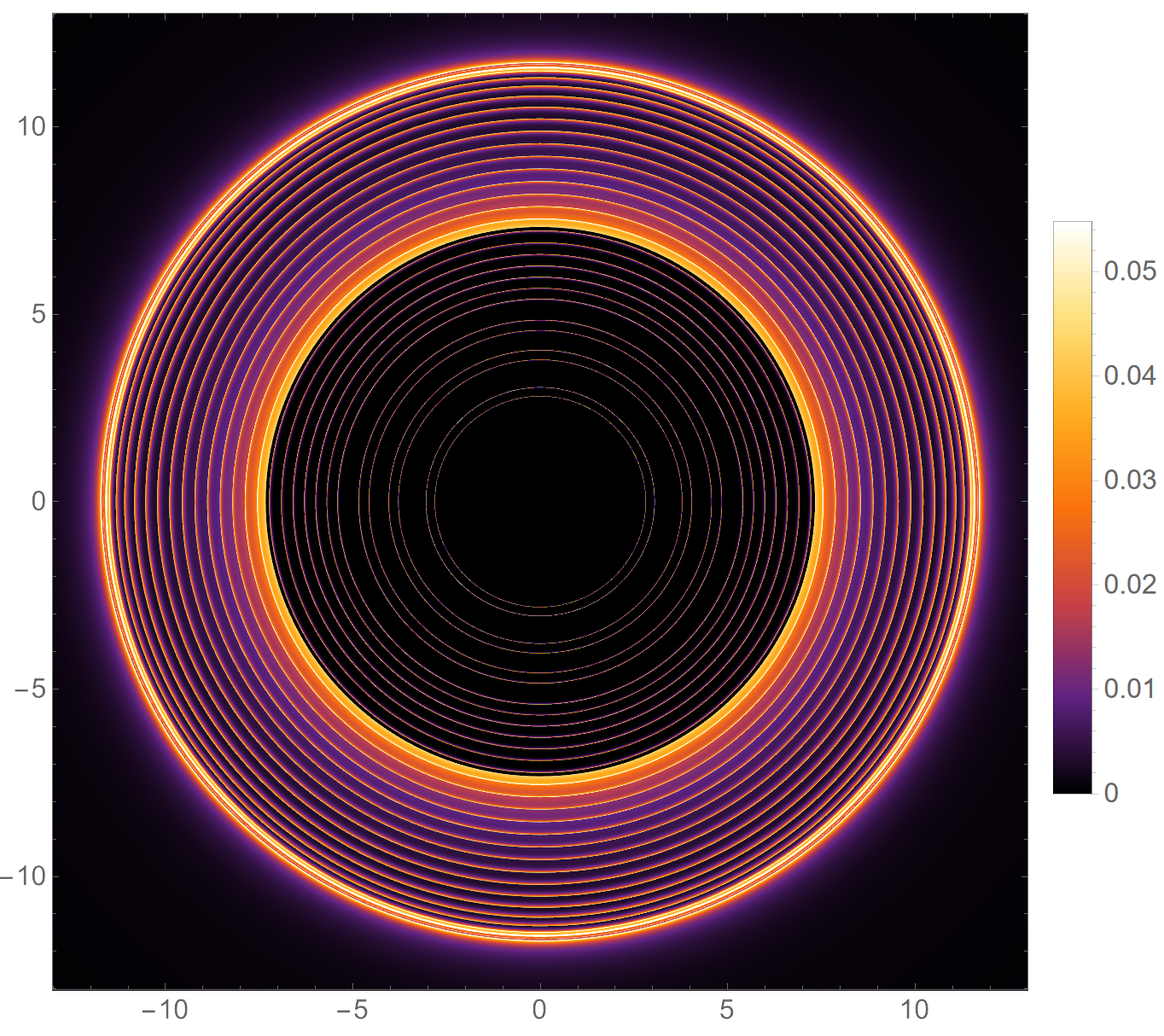}}

\caption{The observed intensity and the appearance for the regular black hole with $b_2=0$ (the first row), $b_2=0.1$ (the second row), $b_2=1$ (the third row), $b_2=10$ (the fourth row) and $b_2=50$ (the fifth row).
The other three parameters of the regular black hole are $r_+=2, M=1, r_-=0.2$.
The first column is the observed intensity when the accretion disk is only located in universe $A$ while the observer is in universe $B$.
And the second column is the corresponding image.
The third column is the observed intensity when the accretion disks are located both in universe $A$ and $B$.
Its image is shown as the fourth column.}
\label{b2}
\end{figure}

Similarly, when we consider the effect of $b_2$ on the observed intensity and the appearance, the corresponding figure is given in Fig.\ref{b2}.
In case $A$, there are more rings as $b_2$ increases.
As can be seen from the fifth row, there is a significant amount and density of the rings when $b_2=50$.
Besides, the radius of the critical curve gets larger as $b_2$ increase.
As for the intensity, they are basically unchanged when $b_2 \in [0,10]$.
But when $b_2$ reaches 50, the intensity drops fifty percent relative to $b_2 \in [0,10]$.
This can be seen more clearly in the figures of appearance.

In case $AB$, the number of rings shows the same pattern as case $A$.
The intensity is a superposition of case $A$ and case $B$.
Thus as $b_2$ increases, there are more rings, the  overall intensity becomes lower, and the radius of the critical curve becomes larger.
The graphs in the right column also illustrate this well.

\section{Conclusions and discussion}\label{S5}
Many people believe that singularities do not exist within the realm of nature.
Therefore, regular black holes which removing the curvature singularity at its center attract a lot of interest.
Although they have no singularities, they usually have a unstable inner horizon accompanied by the mass inflation.
This instability eventually turns the inner horizon into a null singularity which ensures the validity of SCCC.
However, a regular black hole with a stable inner horizon is proposed \cite{Carballo-Rubio:2022kad}.
This inner-extremal regular black hole has a vanishing inner surface gravity, which avoids the mass inflation.
Of course, this is in contradiction with the SCCC and violates the  predictability of classical theories.

In this paper, we study the global causal structure of this regular black hole, and draw its Penrose diagram.
The stable inner horizon allows spacetime to repeat infinitely and makes it possible for the photons to pass through the horizons safely.
Thus the photons from a companion universe can fly to our universe through the black hole-white hole channel.
This process will result in a new appearance of the regular black hole.
Therefore, using a thin accretion disk as the light source, we study the image of this regular black hole by the ray-tracing method.
When the photons are emitted and received at the same universe, the image is similar to that of the Schwarzschild black hole.
Nevertheless, as for the case the accretion disk located around the companion black hole in the preceding universe and the observer located in our universe, the novel multi-ring structure occurs.
There are many extra bright rings in the image, and some of them are distinctly inside the shadow region.
Besides, these rings contain the information about the background geometry, so we may obtain the parameters in the metric from the image.
For this reason, we depict the images with different parameters of the regular black hole and locations of the accretion disk.

As we mentioned before, the regular black hole with a stable Cauchy horizon is in contradiction with the SCCC.
If the SCCC is violated, the multi-ring structure, which is distinctly different from Schwarzschild's, may appear.
However, it is worth noting that such a multi-ring structure can also occur in case of compact objects and wormholes.
With the assistance of other methods that can distinguish between black holes, compact objects, and wormholes, it is possible to test the SCCC in astronomical observations by detecting the multi-ring structure.

Although the regular black hole we studied has a novel image, a precise physical process that leads to the formation of such black hole remains a mystery.
This puts into question whether such a black hole can exist and whether it can be detected in astronomical images.
On the other hand, Cai has found the two-dimensional dilaton black hole and the black hole in higher-derivative gravity theories which have degenerate Cauchy horizons with vanished surface gravities more than twenty years ago~\cite{Cai:1995nt,Cai:1998yp}.
Consequently, their Cauchy horizons are immune from instability.
This provides us with the assurance that we can find a regular black hole with a stable inner horizon in certain theoretical models.
Although, their inner horizons are suggested to exhibit instability when taking into account the quantum effects~\cite{Hollands:2019whz}, whether quantum effects significantly alters the behavior of astronomical massive black holes is still an issue of ongoing debate.
Therefore, the multi-ring structure could serve as a probe into the potential influence of quantum effects on black hole stability.
The emergence of this structure may suggest that quantum effects play a less significant role than previously thought in these massive objects.
This will be the focus of our subsequent research.

\section*{Acknowledgement}

We would like to thank Prof. Rong-Gen Cai for his valuable discussions. This work was supported in part by the National Natural Science Foundation of China with grants No.12075232 and No.12247103.
It is also supported by the National Key R\&D Program of China Grant No.2022YFC2204603.


\begin{thebibliography}{99}

\bibitem{EventHorizonTelescope:2019dse}
K.~Akiyama \textit{et al.} [Event Horizon Telescope],
Astrophys. J. Lett. \textbf{875}, L1 (2019)
doi:10.3847/2041-8213/ab0ec7
[arXiv:1906.11238 [astro-ph.GA]].

\bibitem{EventHorizonTelescope:2022wkp}
K.~Akiyama \textit{et al.} [Event Horizon Telescope],
Astrophys. J. Lett. \textbf{930}, no.2, L12 (2022)
doi:10.3847/2041-8213/ac6674
[arXiv:2311.08680 [astro-ph.HE]].

\bibitem{Gralla:2019xty}
S.~E.~Gralla, D.~E.~Holz and R.~M.~Wald,
Phys. Rev. D \textbf{100}, no.2, 024018 (2019)
doi:10.1103/PhysRevD.100.024018
[arXiv:1906.00873 [astro-ph.HE]].

\bibitem{Penrose:1964wq}
R.~Penrose,
Phys. Rev. Lett. \textbf{14}, 57-59 (1965)
doi:10.1103/PhysRevLett.14.57


\bibitem{Hawking:1970zqf}
S.~W.~Hawking and R.~Penrose,
Proc. Roy. Soc. Lond. A \textbf{314}, 529-548 (1970)
doi:10.1098/rspa.1970.0021

\bibitem{Penrose:1969pc}
R.~Penrose,
Riv. Nuovo Cim. \textbf{1}, 252-276 (1969)
doi:10.1023/A:1016578408204

\bibitem{Bardeen}
J.M.~Bardeen: Non-singular general-relativistic gravitational collapse, Proceedings of the International Conference GR5, Tbilisi, U.S.S.R. (1968), p. 174

\bibitem{Modesto:2008im}
L.~Modesto,
Int. J. Theor. Phys. \textbf{49}, 1649-1683 (2010)
doi:10.1007/s10773-010-0346-x
[arXiv:0811.2196 [gr-qc]].

\bibitem{Hayward:2005gi}
S.~A.~Hayward,
Phys. Rev. Lett. \textbf{96}, 031103 (2006)
doi:10.1103/PhysRevLett.96.031103
[arXiv:gr-qc/0506126 [gr-qc]].

\bibitem{Lewandowski:2022zce}
J.~Lewandowski, Y.~Ma, J.~Yang and C.~Zhang,
Phys. Rev. Lett. \textbf{130}, no.10, 101501 (2023)
doi:10.1103/PhysRevLett.130.101501
[arXiv:2210.02253 [gr-qc]].

\bibitem{Bambi:2023try}
C.~Bambi,
Springer Singapore, 2023,
doi:10.1007/978-981-99-1596-5
[arXiv:2307.13249 [gr-qc]].

\bibitem{Lan:2023cvz}
C.~Lan, H.~Yang, Y.~Guo and Y.~G.~Miao,
[arXiv:2303.11696 [gr-qc]].

\bibitem{Ghosh:2022gka}
R.~Ghosh, M.~Rahman and A.~K.~Mishra,
Eur. Phys. J. C \textbf{83}, no.1, 91 (2023)
doi:10.1140/epjc/s10052-023-11252-0
[arXiv:2209.12291 [gr-qc]].

\bibitem{Carballo-Rubio:2019fnb}
R.~Carballo-Rubio, F.~Di Filippo, S.~Liberati and M.~Visser,
Phys. Rev. D \textbf{101}, 084047 (2020)
doi:10.1103/PhysRevD.101.084047
[arXiv:1911.11200 [gr-qc]].

\bibitem{Brown:2011tv}
E.~G.~Brown, R.~B.~Mann and L.~Modesto,
Phys. Rev. D \textbf{84}, 104041 (2011)
doi:10.1103/PhysRevD.84.104041
[arXiv:1104.3126 [gr-qc]].


\bibitem{Iofa:2022dnc}
M.~Z.~Iofa,
J. Exp. Theor. Phys. \textbf{135}, no.5, 647-654 (2022)
doi:10.1134/S1063776122110048
[arXiv:2206.09460 [gr-qc]].


\bibitem{Carballo-Rubio:2021bpr}
R.~Carballo-Rubio, F.~Di Filippo, S.~Liberati, C.~Pacilio and M.~Visser,
JHEP \textbf{05}, 132 (2021)
doi:10.1007/JHEP05(2021)132
[arXiv:2101.05006 [gr-qc]].

\bibitem{Penrose:1978}
R. Penrose, Singularities of Spacetime, in Theoretical Principles in Astrophysics and Relativity, edited by
N.R. Lebovitz, W. H. Reid, P.O.Vandervoort, 1978

\bibitem{Penrose:1979azm}
R.~Penrose,
``Singularities and time-asymmetry,''
 in General relativity, and Einstein centenary survey, edited
by S.W.Hawking and W. Israel(1979).

\bibitem{Dafermos:2002ka}
M.~Dafermos,
Contemp. Math. \textbf{350}, 99-113 (2004)
[arXiv:gr-qc/0209052 [gr-qc]].

\bibitem{Dafermos:2003vim}
M.~Dafermos,
Ann. Math \textbf{158}, no.3, 875-928 (2003)

\bibitem{Dafermos:2003wr}
M.~Dafermos,
Commun. Pure Appl. Math. \textbf{58}, 0445-0504 (2005)
[arXiv:gr-qc/0307013 [gr-qc]].

\bibitem{Dafermos:2012np}
M.~Dafermos,
Commun. Math. Phys. \textbf{332}, 729-757 (2014)
doi:10.1007/s00220-014-2063-4
[arXiv:1201.1797 [gr-qc]].

\bibitem{Dafermos:2017dbw}
M.~Dafermos and J.~Luk,
[arXiv:1710.01722 [gr-qc]].

\bibitem{Guo:2021wid}
Y.~Guo and Y.~G.~Miao,
Nucl. Phys. B \textbf{983}, 115938 (2022)
doi:10.1016/j.nuclphysb.2022.115938
[arXiv:2112.01747 [gr-qc]].

\bibitem{DeMartino:2023ovj}
I.~De Martino, R.~Della Monica and D.~Rubiera-Garcia,
[arXiv:2310.11039 [gr-qc]].

\bibitem{Tsukamoto:2017fxq}
N.~Tsukamoto,
Phys. Rev. D \textbf{97}, no.6, 064021 (2018)
doi:10.1103/PhysRevD.97.064021
[arXiv:1708.07427 [gr-qc]].

\bibitem{Tsukamoto:2014tja}
N.~Tsukamoto, Z.~Li and C.~Bambi,
JCAP \textbf{06}, 043 (2014)
doi:10.1088/1475-7516/2014/06/043
[arXiv:1403.0371 [gr-qc]].

\bibitem{Jiang:2023img}
H.~X.~Jiang, C.~Liu, I.~K.~Dihingia, Y.~Mizuno, H.~Xu, T.~Zhu and Q.~Wu,
[arXiv:2312.04288 [gr-qc]].

\bibitem{Zhang:2023okw}
C.~Zhang, Y.~Ma and J.~Yang,
Phys. Rev. D \textbf{108}, no.10, 104004 (2023)
doi:10.1103/PhysRevD.108.104004
[arXiv:2302.02800 [gr-qc]].

\bibitem{Cao:2023aco}
L.~M.~Cao, L.~Y.~Li, L.~B.~Wu and Y.~S.~Zhou,
[arXiv:2308.10746 [gr-qc]].


\bibitem{Carballo-Rubio:2022kad}
R.~Carballo-Rubio, F.~Di Filippo, S.~Liberati, C.~Pacilio and M.~Visser,
JHEP \textbf{09}, 118 (2022)
doi:10.1007/JHEP09(2022)118
[arXiv:2205.13556 [gr-qc]].

\bibitem{Olmo:2021piq}
G.~J.~Olmo, D.~Rubiera-Garcia and D.~S.~C.~G\'omez,
Phys. Lett. B \textbf{829}, 137045 (2022)
doi:10.1016/j.physletb.2022.137045
[arXiv:2110.10002 [gr-qc]].

\bibitem{Guerrero:2022qkh}
M.~Guerrero, G.~J.~Olmo, D.~Rubiera-Garcia and D.~G\'omez S\'aez-Chill\'on,
Phys. Rev. D \textbf{105}, no.8, 084057 (2022)
doi:10.1103/PhysRevD.105.084057
[arXiv:2202.03809 [gr-qc]].

\bibitem{Guerrero:2022msp}
M.~Guerrero, G.~J.~Olmo, D.~Rubiera-Garcia and D.~S\'aez-Chill\'on G\'omez,
Phys. Rev. D \textbf{106}, no.4, 044070 (2022)
doi:10.1103/PhysRevD.106.044070
[arXiv:2205.12147 [gr-qc]].

\bibitem{Olmo:2023lil}
G.~J.~Olmo, J.~L.~Rosa, D.~Rubiera-Garcia and D.~Saez-Chillon Gomez,
Class. Quant. Grav. \textbf{40}, no.17, 174002 (2023)
doi:10.1088/1361-6382/aceacd
[arXiv:2302.12064 [gr-qc]].










\bibitem[Chandrasekhar and Hartle(1982)]{1982RSPSA.384..301C}
S.~Chandrasekhar,  and J.~B.~Hartle, : 1982, {\it Proceedings of the Royal Society of London Series A} {\bf 384}, 301. doi:10.1098/rspa.1982.0160.



\bibitem{Ori:1991zz}
A.~Ori,
Phys. Rev. Lett. \textbf{67}, 789-792 (1991)
doi:10.1103/PhysRevLett.67.789

\bibitem{Carter:2009nex}
B.~Carter,
Gen. Rel. Grav. \textbf{41}, no.12, 2873-2938 (2009)
doi:10.1007/s10714-009-0888-5

\bibitem{Walker:1970}
M.~Walker,
J. Math. Phys. \textbf{11}, 2280-2286 (1970)
doi:10.1063/1.1665393









\bibitem{Rybicki:2004hfl}
G.~B.~Rybicki,
Wiley-VCH, 2004,
ISBN 978-0-471-82759-7, 978-3-527-61817-0
doi:10.1002/9783527618170


\bibitem{Whittaker:1988}
E.~Whittaker, S.~McCrae.
A Treatise on the Analytical Dynamics of Particles and Rigid Bodies (Cambridge Mathematical Library), 1988.
Cambridge: Cambridge University Press. doi:10.1017/CBO9780511608797




\bibitem{Cai:1995nt}
R.~G.~Cai,
Phys. Rev. D \textbf{53}, 5698-5704 (1996)
doi:10.1103/PhysRevD.53.5698

\bibitem{Cai:1998yp}
R.~G.~Cai,
Phys. Rev. D \textbf{59}, 104004 (1999)
doi:10.1103/PhysRevD.59.104004
[arXiv:gr-qc/9810071 [gr-qc]].

\bibitem{Hollands:2019whz}
S.~Hollands, R.~M.~Wald and J.~Zahn,
Class. Quant. Grav. \textbf{37}, no.11, 115009 (2020)
doi:10.1088/1361-6382/ab8052
[arXiv:1912.06047 [gr-qc]].

\end{thebibliography}
\end{document}